\documentclass{aa}
\usepackage{graphicx}
\usepackage{txfonts}

\usepackage{subcaption}
\usepackage{lscape}
\usepackage{placeins}

\usepackage[colorlinks=true, linkcolor=blue, citecolor=blue, filecolor=blue, urlcolor=blue]{hyperref}

\usepackage{multirow}
\usepackage{xcolor}
\usepackage{colortbl}

\begin{document}

\title{An empirical view of the extended atmosphere and inner envelope of the asymptotic giant branch star R Doradus}

\subtitle{II. Constraining the dust properties with radiative transfer modelling}

\titlerunning{Dust properties and wind-driving in R Doradus}

\author{Thiebaut Schirmer\inst{1} \and Theo Khouri\inst{1} \and Wouter Vlemmings\inst{1} \and Gunnar Nyman\inst{2} \and Matthias Maercker\inst{1} \and Ramlal Unnikrishnan\inst{1} \and Behzad Bojnordi Arbab\inst{1} \and Kirsten K. Knudsen\inst{1} \and Susanne Aalto\inst{1}}

\authorrunning{T. Schirmer et al.}

\institute{
    Department of Space, Earth and Environment, Chalmers University of Technology, 412 96, Gothenburg, Sweden. \
    \email{thisch@chalmers.se}
    \and
    Department of Chemistry and Molecular Biology, Gothenburg University, Box 462, 40530 Gothenburg, Sweden \
}

\date{Received August 18, 2025; accepted September, 24 2025}

\abstract{
    {Context. Mass loss in oxygen-rich asymptotic giant branch (AGB) stars remains a longstanding puzzle, as the dust species detected around these stars appear too transparent to drive winds through the absorption of radiation alone. The current paradigm consists of outflows driven by photon scattering and requires relatively large grains ($\sim 0.3~\mu$m). Whether the necessary number of grains with the required scattering properties exist around AGB stars remains to be determined empirically.}

   {Aims. We test whether the dust grains observed around the oxygen-rich AGB star R Doradus can drive its stellar wind by combining, for the first time, polarimetric constraints with elemental abundance limits and force balance calculations. We examine Fe-free silicates (MgSiO$_3$), aluminium oxide (Al$_2$O$_3$), and Fe-bearing silicates (MgFeSiO$_4$) to determine whether any dust species can generate sufficient radiative pressure under physically realistic conditions.}

    {Methods. We analysed high-angular-resolution polarimetric observations obtained with SPHERE/ZIMPOL at the Very Large Telescope (VLT) and modelled the circumstellar dust using the radiative transfer code RADMC-3D. Dust optical properties were computed using Optool for both Mie and the distribution of hollow spheres (DHS) scattering theories. By systematically exploring a six-dimensional parameter space, we derived constraints on dust grain sizes, density profiles, and wavelength-dependent stellar radii. For models that successfully fit the observations, we analysed the results taking into consideration recent models for the gas density distribution around R Dor, and applied a multi-criteria zone analysis incorporating gas-depletion constraints and radiation pressure thresholds to assess dust-driven wind viability.}

    {Results. We find sub-micron MgSiO$_3$ and Al$_2$O$_3$ grains (up to 0.1~$\mu$m) regardless of scattering theory considered, and a two-layer dust envelope with steep density profiles (r$^{-3.4}$ to r$^{-4.1}$).
    Despite matching observed scattered-light patterns, these grains generate insufficient radiative force under physically realistic gas-to-dust mass ratios, even when assuming complete elemental depletion.
    Silicates containing Fe could theoretically provide adequate force, but would sublimate in critical acceleration regions and require implausibly high silicon-depletion levels.}

    {Conclusions. Our findings for R Doradus show insufficient radiation pressure from scattering on grains, suggesting that dust alone cannot drive the wind in this star and that additional mechanisms may be required.}
}


\maketitle
\nolinenumbers  

\section{Introduction}
At the final stages of their lives, low- and intermediate-mass stars ascend the asymptotic giant branch (AGB).
In this phase, stars consist of an inert core of carbon and oxygen surrounded by nuclear burning shells and a convective envelope \citep{herwig_evolution_2005}.
The stellar envelope expands dramatically, while pulsations and convection extend the atmosphere, creating conditions suitable for dust condensation.
The current mass-loss paradigm consists of radiation pressure acting on dust grains to overcome the gravitational pull of the star and trigger a slow and dense wind \citep[e.g.][]{hofner_mass_2018}.

However, oxygen-rich AGB stars present a particularly challenging case.
The nucleation of oxygen-rich dust is thought to begin with materials such as aluminium oxide (Al$_2$O$_3$), followed by Fe-free silicate grains (MgSiO$_3$) \citep{karovicova_new_2013, hofner_dynamic_2016}.
A theoretical crisis emerged when it was demonstrated that Fe-free silicates - the dominant species inferred from observations - are insufficiently opaque to drive winds \citep{woitke_too_2006}.
Their work also demonstrated that while Fe-bearing silicates (MgFeSiO$_4$) could theoretically provide sufficient opacity, these grains would become too warm to exist in the inner envelope where wind acceleration begins.
It was instead proposed that Fe-free silicate grains could drive winds if they grow to sizes of at least 0.3 $\mu$m, as larger grains effectively scatter stellar photons despite their low absorption cross-sections \citep{hofner_winds_2008}.

High-angular-resolution observations of polarised light have revolutionised our ability to test these models by directly studying light scattered by dust in the dust-formation and wind-acceleration zones \citep{norris_close_2012, khouri_study_2016, ohnaka_clumpy_2016, khouri_inner_2020}. 
These observations confirm the presence of dust grains roughly in the expected size range around oxygen-rich AGB stars, but whether these observed dust populations can actually drive the winds under physically realistic conditions remains to be determined.

R Doradus (R Dor), the closest oxygen-rich AGB star to the Sun at a distance of $\sim$59 pc \citep{knapp_reprocessing_2003, khouri_empirical_2024}, provides an ideal laboratory for investigating this question.
The combination of its proximity and large angular size (diameter $\sim$62.2 mas; \citep{vlemmings_rotation_2018}) enables spatially resolved observations of the dust-formation and wind-acceleration regions with unprecedented detail, making it particularly suitable for testing theoretical wind-driving models against observational constraints.
R Dor exhibits a semi-regular variable pattern with alternating pulsation modes having periods of 175 and 332 days \citep{bedding_mode_1998}, an effective temperature of $\sim$3000–3100 K \citep{dumm_stellar_1998}, and a moderate mass-loss rate of $\sim10^{-7}$ M$\odot$ yr$^{-1}$ \citep{van_de_sande_chemical_2018}.
High-resolution polarised-light \citep{khouri_study_2016} and molecular-emission \citep{vlemmings_rotation_2018, decin_alma_2018, nhung_morpho-kinematics_2021, khouri_empirical_2024} observations have revealed complex circumstellar features, including rotation and asymmetric structures, which provide crucial constraints for radiative-transfer modelling.

Earlier studies have focused primarily on fitting polarimetric observations without directly testing whether the observed amount and properties of the dust and gas are consistent with the wind-driving paradigm \citep[e.g.][]{khouri_study_2016, ohnaka_clumpy_2016}. 
Here, we analyse the dust in the immediate vicinity of R Dor in detail by systematically exploring a six-dimensional parameter space defined by dust grain size, density profile, and stellar radii, to derive constraints on the dust properties in the inner circumstellar environment.
For the first time, we combine constraints from the dust model with results for the gas-density distribution \citep{khouri_empirical_2024} to test whether such models can satisfy the physical requirements for wind-driving, combining force-balance calculations with elemental-abundance constraints.
Our analysis of R Dor provides a crucial empirical test of the mass-loss mechanism in oxygen-rich AGB stars.

The structure of this paper is as follows. In Section~\ref{sec:observations}, we describe the SPHERE/ZIMPOL observations and data reduction process. 
In Section~\ref{sec:RT_modeling}, we outline the radiative-transfer modelling with RADMC-3D, including the definition of the dust density profile and the parameter space explored, and we present the methodology used to derive the optical properties of dust grains. Section~\ref{sec:model_grid} details the model grid and simulation strategy, describing the iterative approach used to optimise computational efficiency and ensure convergence. The results of our modelling, including constraints on dust grain size, composition, and polarimetric fits, are presented in Section~\ref{sec:results}. In Section~\ref{sec:elemental-budget-limits}, we examine elemental and radiative constraints through a detailed analysis of MgSiO$_3$ models using DHS scattering. Section~\ref{subsec:dust_driven_wind} synthesises the wind-driving potential across all dust compositions and scattering theories. In Section~\ref{sec:discussion}, we compare our findings with previous studies and discuss their implications for AGB mass-loss mechanisms. Finally, Section~\ref{subsec:summary_conclusion} summarises our key results and suggests directions for future work.

\section{Observations and data reduction}
\label{sec:observations}

R Dor was observed using SPHERE/ZIMPOL on the Very Large Telescope on November 28 and 29, 2017 (ESO program 0100.D-0737). 
The observations were carried out with three filters in polarimetric mode, with cycles consisting of blocks of exposures for each of the four angles of the polarimeter. 
Observations of a point spread function (PSF) reference star (HD 25170) were obtained before those of R Dor, following the same procedure but with different exposure times.
Table~\ref{tab:observations} summarises the observational setup and integration times for both the target and the PSF reference. The PSF maps are shown in Appendix~\ref{sec:PSF}.

\begin{table*}[h]
    \centering
    \caption{SPHERE/ZIMPOL observation details}
    \label{tab:observations}
    \begin{tabular}{lcccccc}
        \hline\hline
        Filter & $\lambda_{\rm eff}$ & \multicolumn{2}{c}{R Doradus} & \multicolumn{2}{c}{HD 25170 (PSF)} \\
               & ($\mu$m) & Exp. time & Cycles & Total time & Exp. time & Total time \\
               &          & (s)       &        & (min)      & (s)       & (min) \\
        \hline
        CntHa  & 0.65     & 1.1       & 21     & 30.8       & 14        & 11.2 \\
        Cnt748 & 0.75     & 1.1       & 6      & 7.9        & 1.3       & 10.4 \\
        Cnt820 & 0.82     & 1.1       & 6      & 7.9        & 1.3       & 10.4 \\
        \hline
    \end{tabular}
    \tablefoot{\textbf{Notes.} Observations of R Dor and the PSF reference star (HD 25170). Each cycle consists of blocks with 20 exposures for CntHa and 18 exposures for Cnt748 and Cnt820, covering the four polarimeter angles.}
\end{table*}
 
The data were reduced using the pipeline developed at ETH Zurich \citep{schmid_spherezimpol_2018}, with the standard steps of bias subtraction, flat-field correction, and polarimetry modulation-demodulation efficiency correction. 
We adopted a value of 0.8 for the polarimetric efficiency following \citet{schmid_spherezimpol_2018}. 
The pipeline produces images of the Stokes parameters I (total intensity), Q,  and U. 
We calculated the linearly polarised intensity using $I_{\rm p} = \sqrt{Q^2 + U^2}$, the polarisation degree using $p = I_{\rm p}/I$, and the angle of the polarised light using $\theta = arctan(U/Q)/2$.

\begin{figure*}[h]
\centering
\includegraphics[width=\textwidth, trim={0 0cm 0cm 0cm},clip]{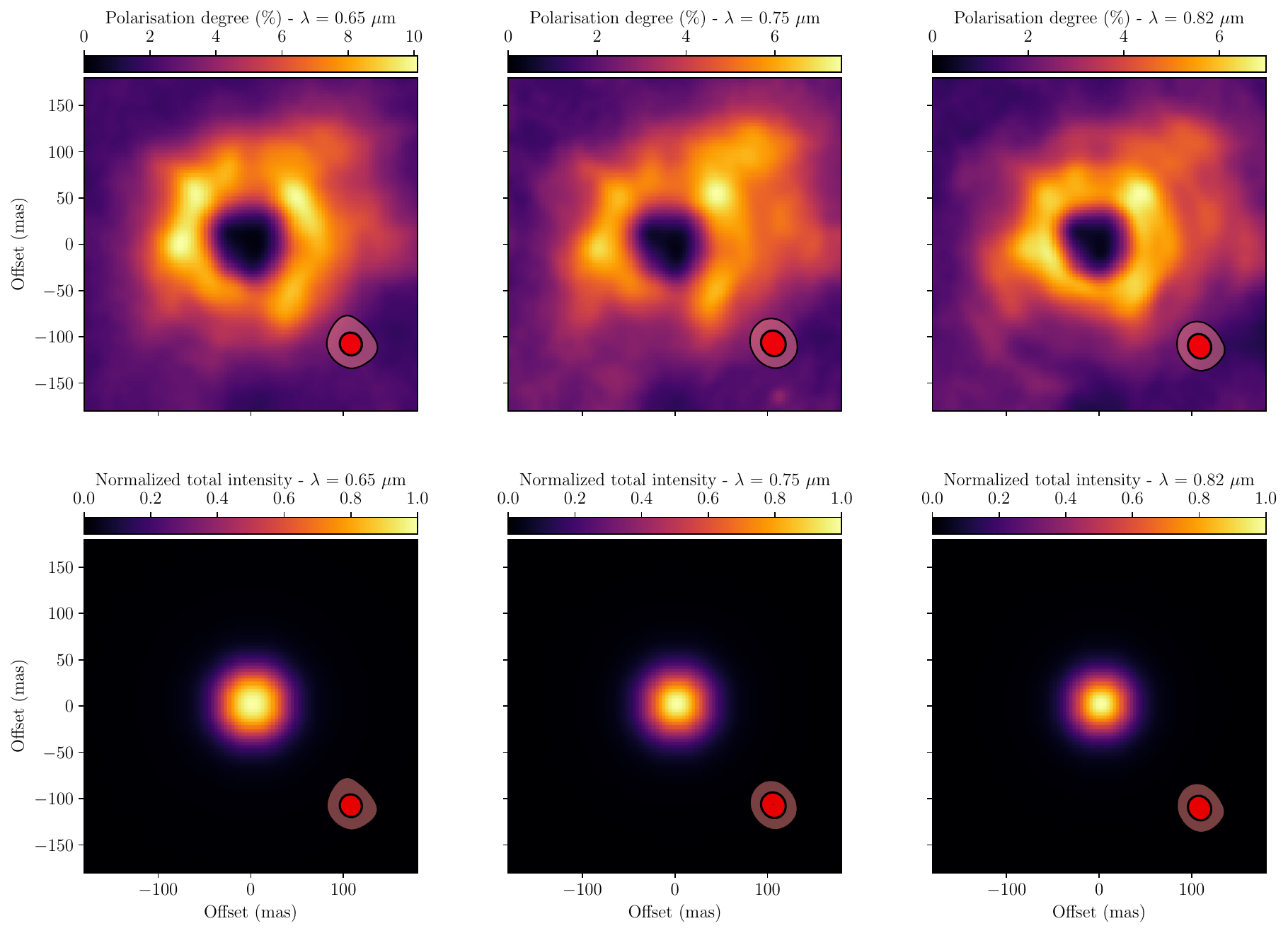}
    \caption{Polarisation degree maps (top row) and normalised total intensity maps (bottom row) of R Dor observed with VLT/SPHERE-ZIMPOL at three wavelengths: 0.65 $\mu$m (left), 0.75 $\mu$m (middle), and 0.82 $\mu$m (right). The red contours in the bottom right of each panel show the instrument PSF at the 10\% and 50\% intensity levels (see Appendix~\ref{sec:PSF} for the complete PSF maps).}
    \label{fig:PD_I_Obs}
\end{figure*}

Figure~\ref{fig:PD_I_Obs} shows the polarisation degree and normalised total intensity maps of R Dor obtained with VLT/SPHERE-ZIMPOL at three wavelengths (0.65, 0.75, and 0.82 \textmu m). 
These observations reveal the unpolarised stellar core surrounded by a bright arc corresponding to stellar light scattered by dust grains in the circumstellar environment of this AGB star.
The scattered-light pattern provides direct evidence for the presence of dust grains in the immediate vicinity of R Dor.

\section{Radiative transfer modelling}
\label{sec:RT_modeling}

\subsection{Dust grains and optical properties}
\label{sec:dust_properties_concise}

\begin{figure*}[h]
    \centering
    \includegraphics[width=\textwidth, trim={0 0cm 0cm 0cm},clip]{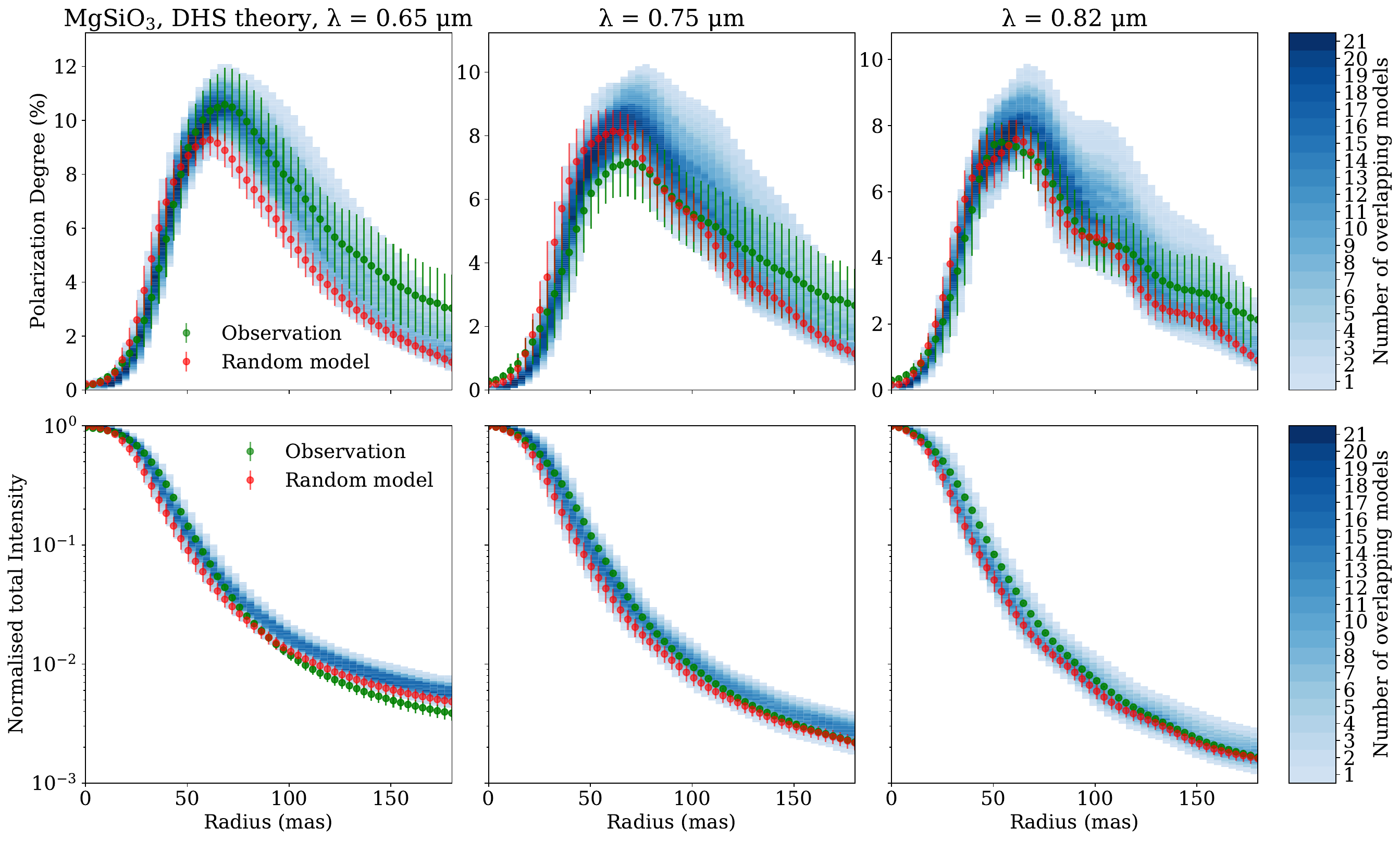}
    \caption{Radial profiles of the polarisation degree (top row) and normalised total intensity (bottom row) for R Dor at three wavelengths: 0.65 \textmu m (left), 0.75 \textmu m (middle), and 0.82 \textmu m (right). The profiles are averaged over 360 degrees. The green curves with error bars represent the observational data. The red curves show one of the acceptable models obtained in Fig.~\ref{fig:chi2min_pyr-mg100_DHS}, while the blue shading indicates the density of overlapping acceptable models.}
        \label{fig:PD_Itot_all_models_pyr-mg100_DHS}
    \end{figure*}

We used the publicly available Optool code \citep{dominik_optool_2021} to compute the optical properties (e.g. scattering phase functions, absorption efficiencies, and Mueller matrices) of dust grains commonly observed around AGB stars. Optool seamlessly interfaces with RADMC-3D, allowing for straightforward adoption of its outputs in radiative transfer simulations. 

Our study considers three main dust compositions: magnesium silicate (MgSiO$_3$), aluminium oxide (Al$_2$O$_3$), and iron-bearing silicate (MgFeSiO$_4$).
Although MgFeSiO$_4$ is prone to overheating in the inner regions of an oxygen-rich AGB star \citep{woitke_too_2006}, we included it for comparison to assess the impact of its stronger near-IR absorption on the wind-driving efficiency. For each material, we computed optical properties using two scattering theories: (i) classical Mie theory for homogeneous, perfectly spherical grains and (ii) the DHS method, which approximates irregular grain shapes and can better reproduce observed polarisation signatures \citep{min_modeling_2005}. 

In addition, we adopted a single dust size to explore the effects of grain size on scattering and absorption.
Although real dust populations follow broader size distributions, constraining the observed polarisation and intensity profiles with a single grain size introduces fewer free parameters.
This single-size approach thus offers a baseline for characterising the dust in R Dor.
The radiative transfer code RADMC-3D \citep{dullemond_radmc-3d_2012} is a versatile tool widely used to model dust and gas in astrophysical environments.
The tool uses Monte Carlo and ray-tracing techniques to simulate radiative processes such as scattering, absorption, and emission on a 3D spatial grid, making it ideal for investigating circumstellar environments like that of R Dor. Importantly, RADMC-3D's ability to compute full Stokes parameters (I, Q, U) makes it particularly well-suited for modelling polarisation and for comparison with SPHERE/VLT observations.

Previous studies \citep[e.g.][]{maercker_investigating_2022, wiegert_asymmetries_2024} have used RADMC-3D for AGB star environments.
Building on this foundation, we applied RADMC-3D to constrain dust grain size, density profiles, and compositions around R Dor. The combination of Optool and RADMC-3D ensured modelling of scattering phase functions and produced polarisation maps, enabling a detailed comparison with observations. The inputs to RADMC-3D were the dust density distribution and the grain optical properties. To consider direction-dependent scattering, the Mueller scattering matrices needed to be provided.

\subsection{Wavelength-dependent stellar radii and dust density profile}

The stellar radius, $R_\lambda$, clearly varies across the three wavelengths of observation (see Figure~\ref{fig:PD_I_Obs}).
We attribute this to changes in optical depth and molecular opacity.
We consider the stellar radii corresponding to the three wavelengths of interest as free parameters: $R_{\lambda_1}$ at $\lambda_1 = 0.65 \, \mu\mathrm{m}$, $R_{\lambda_2}$ at $\lambda_2 = 0.75 \, \mu\mathrm{m}$, and $R_{\lambda_3}$ at $\lambda_3 = 0.82 \, \mu\mathrm{m}$.
Accurately modelling $R_{\lambda}$ is essential for capturing the radiative transfer effects that influence dust scattering and polarisation.

The dust density profile in the inner circumstellar envelope of R Dor was modelled using a power-law distribution

\begin{equation} 
\rho(r) = \rho_0 \times \left(\frac{r}{r_0}\right)^{p_\rho},
\end{equation}

\noindent where $\rho_0$ represents the dust density at the reference radius $r_0 = 180 \, R_\odot$. Fixing $r_0$ allows the normalisation of the dust density to be decoupled from the wavelength-dependent stellar radii $R_{\lambda_1}$, $R_{\lambda_2}$, and $R_{\lambda_3}$. This approach ensures consistency when comparing models across different wavelengths. The power-law exponent, $p_\rho$, governs the steepness of the density gradient.

\subsection{Optimisation of RADMC-3D hyperparameters}

Efficient and accurate modelling with RADMC-3D requires careful selection and optimisation of key hyperparameters to balance computational efficiency and the fidelity of the results.
A critical aspect is the spatial grid resolution, which must be carefully chosen to adequately capture steep gradients in dust density near the star.
Adaptive mesh refinement was not used because preliminary tests showed that a fixed resolution sufficiently captured the dust distribution while minimising computational costs.

For our simulations, we used a computational grid of 112 $\times$ 112 $\times$ 112 cells spanning a physical size of 23.78 $\times$ 23.78 $\times$ 23.78 au.
The final pixel size in the images was determined by an intermediary step, as we employed a 2D grid for the ray-tracing stage of the radiative transfer modelling in RADMC-3D. 
In this step, we used a 112 $\times$ 112 pixel grid, which corresponds to a pixel size of 3.6 mas, matching the native SPHERE/ZIMPOL pixel scale. 
This configuration was optimised through a convergence study to ensure that the chosen resolution did not introduce significant numerical artefacts or biases in the results.

Another vital consideration is the number of photon packages used in the Monte Carlo simulations. 
The number of photon packages was adjusted depending on factors such as optical depth and grain size, ensuring convergence across a wide parameter space. The methodology for determining the appropriate photon count for each model is described in detail in Sect.~\ref{sec:simulation_procedure}.

\section{Model grid and simulation strategy}
\label{sec:model_grid}

\subsection{Overall strategy and parameter space}
\label{sec:overall_strategy}

The simulation workflow for modelling the circumstellar environment of R Dor follows a systematic approach designed to optimise computational efficiency while ensuring physical consistency with observations. Our strategy consists of four main steps: (1) exploring a comprehensive parameter space of dust properties and stellar radii; (2) implementing an iterative radiative transfer procedure with adaptive convergence checks; (3) applying a two-stage filtering process based on total intensity profiles to eliminate unphysical models; and (4) selecting best-fit models through combined polarimetric and intensity constraints. This approach provides a robust framework for constraining the dust grain properties, density profiles, and stellar radii while managing computational resources efficiently.

To systematically explore the parameter space governing dust properties and wind dynamics, we varied key parameters over a well-defined range. Each parameter was chosen to encompass plausible physical values while remaining computationally tractable. The grain size, dust density normalisation, and density profile exponent were sampled logarithmically or linearly, depending on the respective explored range. Similarly, the stellar radii at different wavelengths were varied across a range motivated by observational constraints. The specific parameter ranges and their sampling methods are summarised in Table~\ref{tab:parameter_space}.

\begin{table}[h]
    \centering
    \caption{Parameter space explored in the modelling.}
    \label{tab:parameter_space}
    \begin{tabular}{lccc}
        \hline\hline
        Parameter & Range & Sampling Method & Units \\
        \hline
        $a_0$ & 0.005 -- 1.0 & Logarithmic & $\mu$m \\
        $\rho_0$ & $10^{-18}$ -- $10^{-14}$ & Logarithmic & g/cm$^3$ \\
        $p_\rho$ & -5 -- -2 & Linear & -- \\
        $R_{\lambda_1}$ & 180 -- 450 & Linear & $R_{\odot}$ \\
        $R_{\lambda_2}$ & 180 -- 450 & Linear & $R_{\odot}$ \\
        $R_{\lambda_3}$ & 180 -- 450 & Linear & $R_{\odot}$ \\
        \hline
    \end{tabular}
    \tablefoot{\textbf{Notes.} The number of bins used for each parameter is set to \texttt{nbins} = 15, ensuring accurate coverage of the parameter space while maintaining computational efficiency.}
\end{table}

Exploring such a large parameter space poses significant computational challenges, as the total number of models scales as \(\texttt{nbins}^6\). To address this, we carefully balanced grid resolution and computational efficiency, ensuring sufficient parameter coverage while minimising redundant calculations. Based on preliminary tests, we set the number of bins to 15, which was sufficient to accurately describe the different parameter ranges without being excessively large, thereby avoiding an impractical computation time. Additionally, models with extreme parameter combinations - unlikely to fit the observations - were identified and excluded early in the iterative simulation process through our two-stage filtering approach, significantly reducing unnecessary computations.

\subsection{Observational constraints and diagnostic quantities}
\label{sec:diagnostic_quantities}

To ensure that our models reproduced both the polarimetric signatures and the overall intensity distribution of R Dor, we employed two complementary sets of diagnostic quantities. These metrics allowed us to quantitatively assess model performance and implement efficient filtering strategies throughout the simulation process.

\subsubsection{Total intensity profile constraints}

Scattering of radiation on dust grains has been identified as an important source of opacity affecting the sizes of evolved stars at visible wavelengths \citep{ireland_multiwavelength_2004}. Additionally, molecular opacity effects, particularly from TiO absorption, contribute to wavelength-dependent variations in stellar radii, although these effects are generally less dominant than dust scattering in the circumstellar regions  investigated here \citep{ireland_multiwavelength_2004}. The amount of dust required to fit the observed amount of polarised light in R Dor is sufficient to significantly affect the stellar size at the observed wavelengths. Hence, it was necessary to verify whether the models that provided acceptable fits to the observed polarised-light images were consistent with the total-intensity images.

To compare the modelled and observed total intensity profiles, we analysed their normalised profiles by measuring the full widths at various fractions of the maximum intensity. These fractions represent specific intensity levels: 50\% of the peak normalised intensity, defined as the full width at half maximum (FWHM), 10\%, and 1\%. Measuring full widths at different intensity levels provides complementary information: the FWHM traces the compact core of the stellar disc, while the full widths at 10\% and 1\% probe the extended scattered light distribution. This approach ensured that the focus remained on the overall shape of the intensity profiles.

To quantify the agreement between the modelled and observed intensity profiles, we defined the following diagnostic parameter:

\begin{equation}
    \Delta\mathrm{FW}_j(\%) = \mathrm{FW}_{\mathrm{mod}, j}(\%) - \mathrm{FW}_{\mathrm{obs}, j}(\%),
\end{equation}

\noindent where $\mathrm{FW}_{\mathrm{mod}, j}(\%)$ and $\mathrm{FW}_{\mathrm{obs}, j}(\%)$ are the full widths at the specified intensity fraction for the modelled and observed profiles, respectively, at wavelength $j$. For example, $\Delta\mathrm{FW}_3(50\%)$ represents the difference in FWHM at the longest wavelength ($\lambda_3 = 0.82 \, \mu\mathrm{m}$), while $\Delta\mathrm{FW}_1(1\%)$ corresponds to the difference at 1\% of the maximum intensity for the shortest wavelength ($\lambda_1 = 0.65 \, \mu\mathrm{m}$).

This diagnostic parameter served two critical functions in our analysis. First, it provided an early filtering criterion during the iterative simulation procedure: models with $\Delta\mathrm{FW}_j(50\%) > 200\%$ (i.e. FWHM more than three times the observed value) at any wavelength were immediately excluded as these represented highly optically thick models incompatible with observations. Second, it served as a final selection criterion, where we required $|\Delta\mathrm{FW}_j(\%)|< 30\%$ for all intensity levels and wavelengths. This tolerance threshold accounted for possible gas-emission contributions to the observed profiles, while ensuring that the modelled dust profiles reproduced the overall shape and spatial extent of the observed intensity distribution.

Gas emission and absorption can affect total intensity profiles, particularly in the outer circumstellar regions. Because the radiative transfer simulations focus solely on dust scattering and absorption, they do not account for thermal emission from the gas. This is a shortcoming of the current state-of-the-art modelling of these types of observations, but we expected a relatively small effect in the region where polarised light is observed. The 30\% tolerance threshold was chosen to account for these effects while still maintaining meaningful constraints on the dust distribution.

\subsubsection{Polarisation degree constraints}

The polarisation degree profile as a function of radius served as the primary observable to constrain dust properties. We quantified the comparison between the modelled and observed polarisation profiles using chi-square analysis, with the detailed formulation presented in Section~\ref{sec:results}. We required the models to achieve $\chi^2_{\mathrm{red, 3D}} < 1.36$ (corresponding to the 3$\sigma$ confidence level) for them to qualify as acceptable fits to the polarimetric data. This constraint ensured that the selected models accurately reproduced the observed polarisation signatures across all three wavelengths.

\subsection{Iterative simulation and model-selection procedure}
\label{sec:simulation_procedure}

The simulation procedure implemented an adaptive approach that balanced computational efficiency with convergence requirements. The workflow integrated early model filtering, iterative convergence checks, and systematic validation against observational constraints.

\subsubsection{Radiative transfer setup and initialisation}

The process began with the set of input parameters, including the dust grain size ($a_0$), density normalisation ($\rho_0$), density profile exponent ($p_\rho$), and stellar radii ($R_{\lambda_1}$, $R_{\lambda_2}$, $R_{\lambda_3}$). Using these parameters, we generated the necessary input files for RADMC-3D, including the dust density profile and optical properties computed with \texttt{Optool}. We initialised the simulation with 20,000 photon packages, and the radiative transfer calculations were performed to generate maps of the Stokes parameters I, Q, and U.

\subsubsection{Convergence assessment}

To ensure a meaningful comparison with observations and assess Monte Carlo convergence, we implemented a systematic convergence check procedure. After each radiative transfer calculation, the modelled 2D intensity and polarisation-degree (PD) maps were first convolved with synthetic symmetrical PSFs. The synthetic PSF was constructed by computing the azimuthal average of the observed PSF at each radius and applying this averaged value uniformly across all azimuthal angles for that radius. This approach ensured that any azimuthal variations in the modelled maps at a fixed radius were purely due to Monte Carlo noise and not instrumental effects, as real PSF azimuthal variations could otherwise be misinterpreted as simulation noise.

The stability of the Monte Carlo simulations was evaluated by monitoring the standard deviation in the intensity and polarisation degree maps over successive iterations. We analysed the 2D total-intensity and PD maps convolved with the synthetic PSF by computing the standard deviation of values over azimuthal angles at each radius. If the standard deviation remained below 5\% of the average value at all radii, the simulation was considered to be converged. If this threshold was not met, the number of photon packages was doubled, and the simulation was repeated. This iterative refinement continued for a maximum of 12 iterations, though post-analysis confirmed that convergence was always achieved within fewer steps.

\subsubsection{Two-stage filtering process}

To optimise computational resources and ensure physical consistency, we implemented a two-stage filtering approach. In the first stage, following the initial convergence assessment, we applied an immediate filter based on the total intensity profile width. If $\Delta\mathrm{FW}_j(50\%) > 200\%$ (i.e. the modelled FWHM was more than three times the observed value) at any of the three wavelengths, the model was excluded from further computation. These models were typically highly optically thick, producing stellar radii significantly larger than observed, making it impossible to match the observed intensity profile. This pre-selection prevented the calculation of computationally intensive models that could not reproduce the observations.

In the second stage, for models that passed the early filtering and achieved convergence, a final validation was performed during the best-fit model selection process (detailed in Section~\ref{sect:Selection_of_Best-Fit_Models}). At this stage, we required $|\Delta\mathrm{FW}_j(\%)|< 30\%$ for all intensity levels (50\%, 10\%, and 1\%) and for all wavelengths. Only models satisfying both this intensity constraint and the polarisation chi-squared criterion were retained as viable solutions.

\subsubsection{Final processing and output}

Once convergence was achieved and the model passed the early filtering stage, the unconvolved intensity and PD maps were convolved with the real, asymmetric instrument PSFs, producing the final simulated maps that were directly comparable to observations. This final convolution step ensured that all instrumental effects were properly accounted for in the model-observation comparison.

The complete iterative simulation procedure provides a systematic and efficient framework for exploring the parameter space while managing computational time\footnote{The total computation time for the iterative simulation procedure is approximately 2-3 days.}. By integrating early optical depth filtering, adaptive convergence checks, and multi-parameter validation, the procedure maximises efficiency and ensures that the final models remain physically realistic and consistent with observational constraints. Moreover, it provides a measure of the uncertainty in the models introduced by the fluctuations inherent to Monte Carlo simulations.

\begin{figure*}[h]
    \centering
    \includegraphics[width=\textwidth, trim={0 0cm 0cm 0cm},clip]{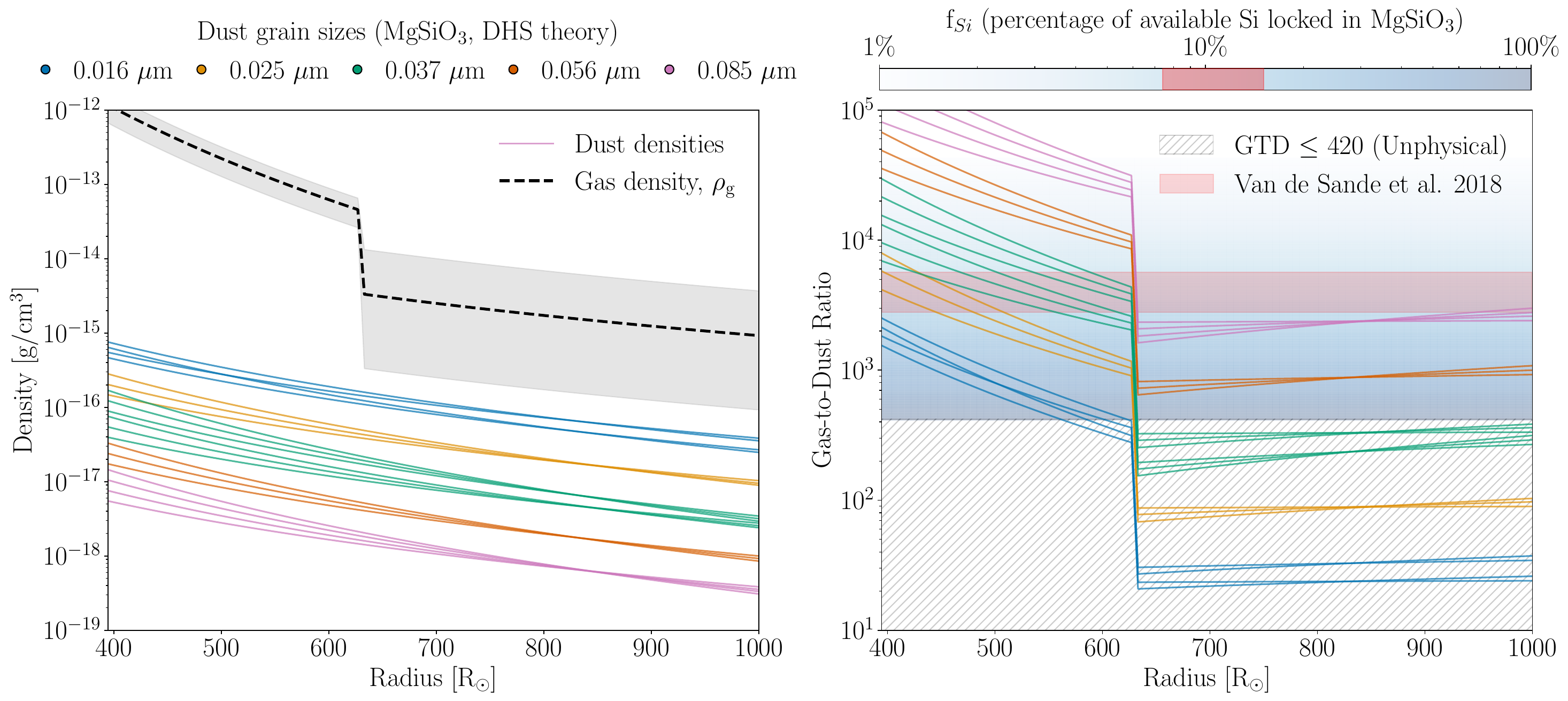}
    \caption{Dust and gas properties around R Dor considering MgSiO$_3$ dust using DHS scattering theory. Left panel: Density profiles showing the gas density (dashed black  line with grey uncertainty region) from \cite{khouri_empirical_2024} and dust densities for different grain sizes (coloured lines) corresponding to models that fit the observational constraints. The gas density exhibits a transition at $1.6 \times R_{887}$ \citep{khouri_empirical_2024}. Right panel: Gas-to-dust (GTD) ratios derived from the density profiles. The hatched area (GTD $\leq$ 420) represents unrealistic values where more silicon would be locked in dust than expected in the stellar atmosphere. The red horizontal band shows constraints from SiO-depletion measurements \citep{van_de_sande_chemical_2018}. The background colour gradient indicates the percentage of available silicon locked in dust grains, from 1\% (light blue) to 100\% (dark blue).}
    
    \label{fig:dust_gas_density_gtd_pyr-mg100_DHS}
    \end{figure*}

\section{Results}
\label{sec:results}

This section presents a systematic exploration of the parameter space, the constraints derived from polarimetric fitting, and the critical wind-driving analysis.

\subsection{Chi-squared analysis}

The goodness of fit between the modelled and observed polarisation degree profiles is quantified using the chi-squared (\(\chi^2\)) statistic, which is calculated as

\begin{equation}
    \chi^2(a_0, \rho_0, p_\rho, R_1, R_2, R_3) = 
    \sum_{j=1}^{3} \sum_{i=1}^{50} \frac{\left(\langle PD_{\mathrm{obs}, i, j} \rangle - \langle PD_{\mathrm{mod}, i, j} \rangle \right)^2}{\sigma_{\mathrm{obs}, i, j}^2 + \sigma_{\mathrm{mod}, i, j}^2},
\end{equation}

\noindent where \( \langle PD_{\mathrm{obs}, i, j} \rangle \) and \( \langle PD_{\mathrm{mod}, i, j} \rangle \) are the mean observed and modelled polarisation degrees at the \(i\)-th radial bin for the \(j\)-th wavelength, respectively. The uncertainties \(\sigma_{\mathrm{obs}, i, j}\) and \(\sigma_{\mathrm{mod}, i, j}\) represent the standard deviations of the observed and modelled polarisation degrees at each radial bin. The summation was performed over 50 radial bins, corresponding to a maximum radius of 180 mas, given the observational pixel size of 3.6 mas. Beyond this radius, the uncertainties in the observed polarisation degree become prohibitively large, with the standard deviation exceeding the polarisation degree itself.

The mean polarisation degree at a given radius, \(r_i\), and wavelength, \(\lambda_j\), is computed as

\begin{equation}
    \langle PD_{i, j} \rangle = \frac{1}{N_{\theta}} \sum_{k=1}^{N_{\theta}} PD_{i, j, k},
\end{equation}

where \( PD_{i, j, k} \) represents the polarisation degree at radius \( r_i \), wavelength \( \lambda_j \), and angular position \( \theta_k \), and \( N_{\theta} \) is the number of angular bins. This expression applies to both the observed (\( PD_{\mathrm{obs}} \)) and modelled (\( PD_{\mathrm{mod}} \)) data.

The standard deviation of the polarisation degree at a given radius and wavelength, accounting for azimuthal variations, is given by

\begin{equation}
    \sigma_{i, j} = \sqrt{\frac{1}{N_{\theta} - 1} \sum_{k=1}^{N_{\theta}} \left( PD_{i, j, k} - \langle PD_{i, j} \rangle \right)^2 }.
\end{equation}

This formulation applies separately to both the observed (\(\sigma_{\mathrm{obs}, i, j}\)) and modelled (\(\sigma_{\mathrm{mod}, i, j}\)) datasets. The reduced chi-square ($\chi^2_{\mathrm{red}}$) is then calculated to account for the degrees of freedom in the fit:

\begin{equation}
    \chi^2_{\mathrm{red}}(a_0, \rho_0, p_\rho, R_{\lambda_1}, R_{\lambda_2}, R_{\lambda_3}) = 
    \frac{\chi^2(a_0, \rho_0, p_\rho, R_{\lambda_1}, R_{\lambda_2}, R_{\lambda_3})}{N - P},
\end{equation}

where $N = 150$ is the total number of data points (50 radial bins for each of the three wavelengths) and $P = 6$ is the number of free parameters ($a_0$, $\rho_0$, $p_\rho$, $R_{\lambda_1}$, $R_{\lambda_2}$, and $R_{\lambda_3}$). 
To remove the dependence of $\chi^2_{\mathrm{red}}$ on the stellar radii ($R_{\lambda_1}$, $R_{\lambda_2}$, $R_{\lambda_3}$), we define a new metric, $\chi^2_{\mathrm{red, 3D}}$, which minimises $\chi^2_{\mathrm{red}}$ over the three radii and the same values of $a_0$, $\rho_0$, and $p_\rho$:

\begin{equation}
    \chi^2_{\mathrm{red, 3D}}(a_0, \rho_0, p_\rho) = 
    \min_{R_{\lambda_1}, R_{\lambda_2}, R_{\lambda_3}} \chi^2_{\mathrm{red}}(a_0, \rho_0, p_\rho, R_{\lambda_1}, R_{\lambda_2}, R_{\lambda_3}).
\end{equation}

The models are classified based on the value of $\chi^2_{\mathrm{red, 3D}}$, with thresholds for 1$\sigma$, 2$\sigma$, and 3$\sigma$ confidence levels set at 1.05, 1.21, and 1.36, respectively. These thresholds, derived using the chi-squared cumulative distribution function, represent acceptable deviations between the model and observations. In the following analysis, we only consider models that fall within the 3$\sigma$ confidence level, meaning that only models with $\chi^2_{\mathrm{red, 3D}} < 1.36$ are retained.

\subsection{Selection of best-fit models}
\label{sect:Selection_of_Best-Fit_Models}

\begin{figure}[h]
\centering
\includegraphics[width=0.5\textwidth, trim={0 0cm 0cm 0cm},clip]{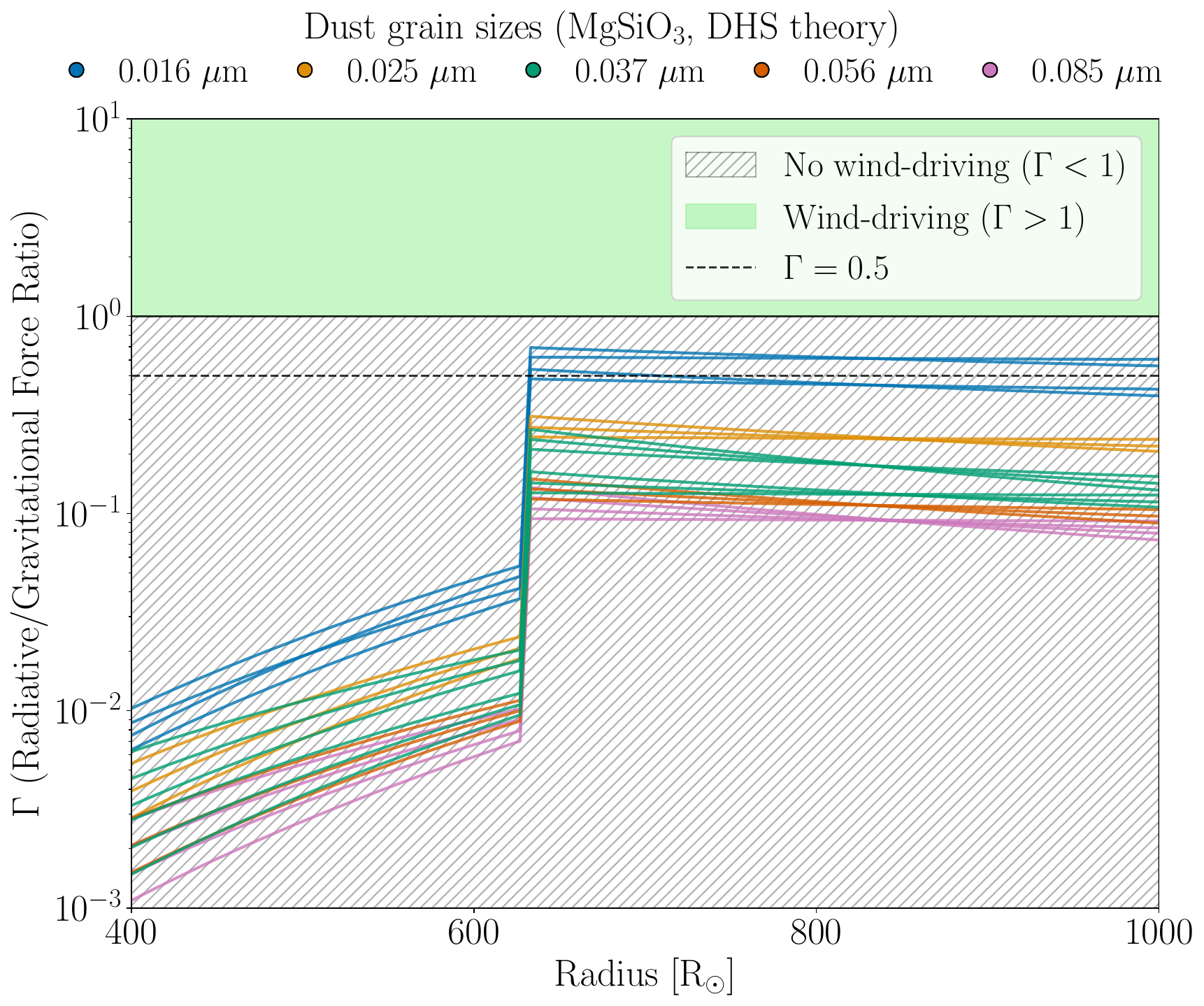}
\caption{Radiative-to-gravitational force ratio ($\Gamma$) as a function of radius for MgSiO$_3$ dust grains calculated using DHS scattering theory. Different coloured lines represent dust grain sizes. The green shaded area ($\Gamma > 1$) indicates regions where radiation pressure exceeds gravity. The hatched area ($\Gamma < 1$) represents regions where radiation pressure alone is insufficient to overcome gravity.}
\label{fig:gamma_values_pyr-mg100_DHS}
\end{figure}

\begin{figure*}[h]
\centering
\includegraphics[width=\textwidth, trim={0 0cm 0cm 0cm},clip]{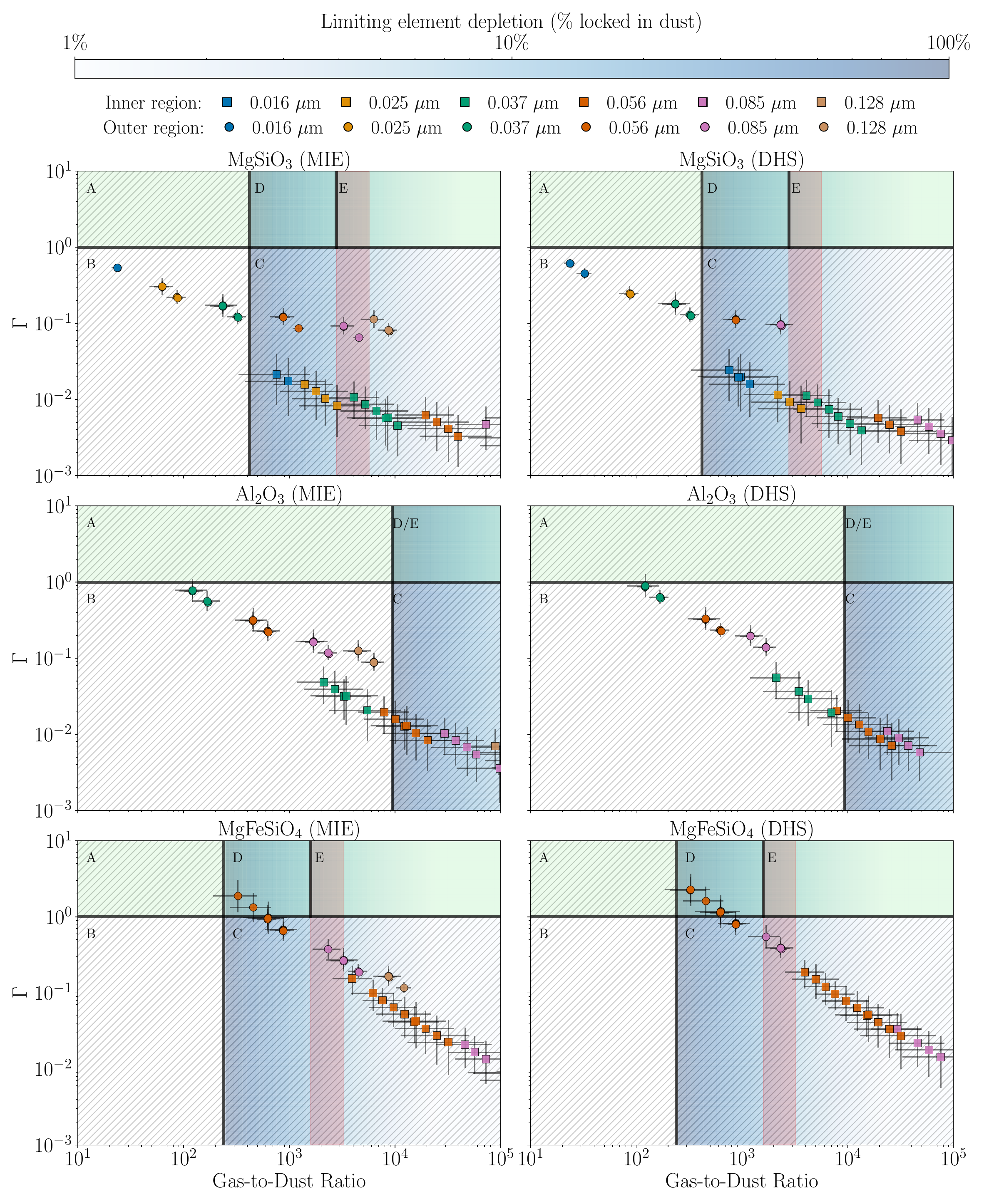}
\caption{Radiative-to-gravitational force ratio ($\Gamma$) vs gas-to-dust mass ratio for dust models fitting the polarisation observations. Squares and circles denote inner and outer regions ($R \lessgtr 1.6 R_{887}$ from \citealt{khouri_empirical_2024}); colours indicate grain sizes. Five zones are shown, based on physical viability: (A) $\Gamma > 1$ but $>100\%$ elemental depletion is impossible; (B) $\Gamma < 1$ and $>100\%$ and depletion is impossible; (C) $<100\%$ depletion but $\Gamma < 1$, representing insufficient radiation pressure; (D) $\Gamma > 1$ with high but possible depletion inconsistent with observations; and (E) $\Gamma > 1$ with observed depletion levels, the ideal wind-driving zone. The horizontal line marks $\Gamma = 1$.}

\label{fig:gtd_gamma_combined_regions}
\end{figure*}

The selection of best-fit models was carried out in two steps: first, by applying thresholds based on the reduced chi-squared statistic (\(\chi^2_{\mathrm{red, 3D}}\)) and second, by filtering models based on the normalised intensity profile differences (\(\Delta\mathrm{FW}_j(\%)\)).

We began by selecting all models with \(\chi^2_{\mathrm{red, 3D}} < 1.36\), ensuring that they provide an acceptable fit to the observed polarisation data.

Among the models that pass this chi-squared threshold, we further refined the selection by comparing the modelled and observed normalised intensity profiles. This comparison was performed using the \(\Delta\mathrm{FW}_j(\%)\) parameters, which quantify the differences in full widths at 50\%, 10\%, and 1\% of the maximum intensity for each wavelength (\(j = 1, 2, 3\)). Specifically, we retained only models for which

\begin{equation}
    |\Delta\mathrm{FW}_j(50\%)|, \, |\Delta\mathrm{FW}_j(10\%)|, \, |\Delta\mathrm{FW}_j(1\%)| < \Delta\mathrm{FW}_{\mathrm{limit}},
\end{equation}

\noindent where \(\Delta\mathrm{FW}_{\mathrm{limit}} = 30\%\). We find that using a smaller limit barely affects the mean, minimum, and maximum values of the parameters we aim to constrain (\(a_0\), \(\rho_0\), \(p_\rho\), \(R_{\lambda_1}\), \(R_{\lambda_2}\), and \(R_{\lambda_3}\)). However, we adopted a 30\% limit to ensure that a sufficient number of models satisfied the criterion for statistical analysis while still maintaining a close match between the modelled and observed intensity profiles.

By combining these two selection criteria, we ensured that the best-fit models provided a comprehensive match to both the polarisation and intensity data.
This approach acknowledges that gas emission may contribute to the observed total intensity profiles while our dust-only radiative transfer models focus on reproducing the polarisation signatures. 
The use of \(\Delta\mathrm{FW}_j(\%)\) as an additional filtering parameter strengthens the reliability of the model selection process by addressing discrepancies in intensity profile shapes across multiple wavelengths. 
In particular, models that provided an acceptable fit to the observed polarisation degree but presented broader total intensity profiles than observed were clearly incompatible with the observations.

Table~\ref{tab:dust_parameter_results} presents the constrained parameter ranges derived from our comprehensive modelling analysis for the three dust compositions using both DHS and Mie scattering theories. 
These results represent the ensemble of models that successfully reproduce the observed polarimetric signatures while maintaining physical consistency with the intensity profiles. 
Detailed analysis of individual best-fit models, including their specific parameter combinations and goodness-of-fit statistics, are provided in Appendix~\ref{sec:best-fit_models}. 
There is a remarkable consistency between Mie and DHS results for dust grain size, dust density, and stellar radii parameters, which strengthens our conclusion that the constraints we derive on these physical parameters are not significantly dependent on assumptions about grain morphology.
As an illustrative example, Figure~\ref{fig:PD_Itot_all_models_pyr-mg100_DHS} demonstrates the excellent agreement between our best MgSiO$_3$ DHS model and observations across the three wavelengths. 
Similar comparisons for the other dust materials and scattering theories are presented in Appendix~\ref{sec:additional_figures}.

\begin{table}
    \caption{Constrained parameter ranges for circumstellar dust around R Dor.}
    \label{tab:dust_parameter_results}
    \centering
    \renewcommand{\arraystretch}{1.3}
    \setlength{\tabcolsep}{10pt}
    \begin{tabular}{lccc}
    \hline\hline
    Parameter & DHS (Min--Max) & Mie (Min--Max) \\
    \hline
    \multicolumn{3}{c}{MgSiO$_3$ dust} \\
    \hline
    $a_0$ [$\mu$m] & 0.02--0.09 & 0.02--0.12 \\
    $\rho_0$ [g/cm$^3$] & $4 \times 10^{-17}$--$10^{-14}$ & $10^{-17}$--$10^{-14}$ \\
    $p_\rho$ & $-4.4$ to $-2.8$ & $-4.4$ to $-2.8$ \\
    $R_{\lambda_1}$ [$R_\odot$] & 288--414 & 306--414 \\
    $R_{\lambda_2}$ [$R_\odot$] & 270--414 & 270--414 \\
    $R_{\lambda_3}$ [$R_\odot$] & 216--378 & 216--378 \\
    \hline
    \multicolumn{3}{c}{MgFeSiO$_4$ dust} \\
    \hline
    $a_0$ [$\mu$m] & 0.06--0.09 & 0.06--0.12 \\
    $\rho_0$ [g/cm$^3$] & $2 \times 10^{-17}$--$10^{-14}$ & $2 \times 10^{-17}$--$10^{-14}$ \\
    $p_\rho$ & $-4.8$ to $-3.0$ & $-4.4$ to $-3.0$ \\
    $R_{\lambda_1}$ [$R_\odot$] & 234--432 & 306--432 \\
    $R_{\lambda_2}$ [$R_\odot$] & 198--432 & 198--432 \\
    $R_{\lambda_3}$ [$R_\odot$] & 198--360 & 198--360 \\
    \hline
    \multicolumn{3}{c}{Al$_2$O$_3$ dust} \\
    \hline
    $a_0$ [$\mu$m] & 0.04--0.09 & 0.04--0.12 \\
    $\rho_0$ [g/cm$^3$] & $6.3 \times 10^{-18}$--$10^{-14}$ & $10^{-17}$--$10^{-14}$ \\
    $p_\rho$ & $-4.8$ to $-3.0$ & $-4.8$ to $-3.0$ \\
    $R_{\lambda_1}$ [$R_\odot$] & 306--414 & 306--432 \\
    $R_{\lambda_2}$ [$R_\odot$] & 216--396 & 198--414 \\
    $R_{\lambda_3}$ [$R_\odot$] & 198--360 & 216--360 \\
    \hline
    \end{tabular}
    \tablefoot{\textbf{Notes.} Models are computed with three different dust compositions (MgSiO$_3$, MgFeSiO$_4$, and Al$_2$O$_3$) using both DHS and Mie scattering theories. Each range represents the minimum and maximum values from models that successfully reproduce the observed polarised-light measurements. All values represent equally valid solutions that successfully reproduce the observed polarised light. Parameters $R_{\lambda_1}$, $R_{\lambda_2}$, and $R_{\lambda_3}$ correspond to stellar radii at wavelengths 0.65, 0.75, and 0.82 $\mu$m.}
    \end{table}

Having successfully identified dust configurations that reproduce the observed polarimetric signatures of R Dor, we now address the fundamental challenge that motivates this study by evaluating whether these observationally constrained dust populations can drive the stellar wind under physically realistic conditions. 
Our assessment incorporates two essential criteria that any viable wind-driving model must satisfy: (1) consistency with elemental gas depletion constraints derived from solar abundances and gas-phase observations, ensuring that the required dust masses do not exceed the available elemental budget; and (2) sufficient radiative pressure to overcome gravitational forces, quantified through the force balance parameter $\Gamma$.

\section{Elemental and radiative constraints: MgSiO$_3$ analysis with DHS}
\label{sec:elemental-budget-limits}
    
Using solar photospheric elemental abundances, we derived (see Appendix~\ref{sec:elemental-abundance-constraints}) absolute lower limits to the gas-to-dust mass ratio, considering full depletion of the limiting element (silicon for silicates and aluminium for alumina).
Assuming complete depletion of these elements, we find minimum gas-to-dust mass ratios of 420 for MgSiO$_3$, 240 for MgFeSiO$_4$, and 9400 for Al$_2$O$_3$. Observations of SiO line emission from R Dor indicate that only $8$-$15$\% of the silicon is actually depleted \citep{van_de_sande_chemical_2018}, raising the realistic limits for the gas-to-dust mass ratios to 2800 for MgSiO$_3$ and 1600 for MgFeSiO$_4$. For Al$_2$O$_3$, no observational constraints on aluminium depletion are available from \citet{van_de_sande_chemical_2018}, so we can only apply the theoretical complete depletion limit. 
We now examine whether our dust models that successfully fit the polarimetric observations can satisfy these physical constraints and drive stellar winds.

Figure~\ref{fig:dust_gas_density_gtd_pyr-mg100_DHS} compares our observationally constrained dust density profiles considering DHS MgSiO$_3$ grains with gas density profiles from CO observations \citep{khouri_empirical_2024}.
Following \citet{khouri_empirical_2024}, we defined two regions based on the gas density profile transition: the inner region ($R < 1.6 \times R_{887}$) and the outer region ($R > 1.6 \times R_{887}$), where $R_{887}$ is the stellar radius at 887 $\mu$m as defined in \citet{khouri_empirical_2024}.
The analysis reveals stark differences between these regions.
In the inner region, models with larger grain sizes (0.037, 0.056, and 0.085 $\mu$m) satisfy observational constraints, requiring silicon depletion levels consistent with or below the 8-15\% \citep[][]{van_de_sande_chemical_2018}.
However, the outer region presents a more restrictive scenario: none of our models align with the observed depletion constraints, as all would require silicon depletion levels exceeding the observationally determined values.

Having established which dust models satisfy elemental abundance constraints, we now examine their wind-driving capability.
The ratio $\Gamma$ of radiative to gravitational acceleration (see Appendix~\ref{sec:force_balance}) serves as the critical test for dust-driven winds, with $\Gamma > 1$ indicating sufficient radiative pressure to overcome gravity.

Figure~\ref{fig:gamma_values_pyr-mg100_DHS} illustrates $\Gamma$ as a function of radius for MgSiO$_3$ dust grains calculated using the DHS approximation. 
The results are unambiguous: regardless of grain size or region, no models achieve $\Gamma > 1$, demonstrating that radiative pressure on dust cannot overcome gravitational forces.
Specifically, models that satisfy the elemental abundance constraints in the inner region do not generate sufficient radiative pressure for wind acceleration.

This analysis for MgSiO$_3$ using the DHS theory reveals a fundamental disconnect: while some dust models are consistent with elemental abundance constraints, none accounts for the expected radiation pressure to drive the observed stellar wind.
Similar comprehensive analyses for all other dust compositions and scattering theories are synthesised in the following section to provide a complete assessment of wind-driving potential across all dust species.

\section{Synthesis of wind-driving potential across dust compositions and scattering theories}
\label{subsec:dust_driven_wind}

We now synthesise results from all dust compositions and scattering theories by applying the two essential criteria established in the previous section.
Figure~\ref{fig:gtd_gamma_combined_regions} shows $\Gamma$ as a function of the gas-to-dust mass ratio for all models in both inner (squares) and outer (circles) envelope regions across the three dust materials and two scattering theories.

We divided the parameter space into five distinct zones:
(A) $\Gamma > 1$ but gas-to-dust mass ratios requiring excessive depletion ($>100\%$ elemental depletion);
(B) $\Gamma < 1$ and gas-to-dust mass ratios requiring excessive depletion ($>100\%$ elemental depletion);
(C) Gas-to-dust ratios requiring $<100\%$ elemental depletion but $\Gamma < 1$;
(D) $\Gamma > 1$ and gas-to-dust mass ratios requiring elemental depletion between observed levels and $100\%$ (depletion potentially viable but inconsistent with observations);
(E) $\Gamma > 1$ and gas-to-dust mass ratios consistent with observed depletion levels (ideal wind-driving zone). 

Across all dust compositions and scattering theories, we find that no models fall within zone E, indicating that none can drive the outflow of R Dor under observationally realistic conditions.

Both MgSiO$_3$ and Al$_2$O$_3$ exhibit similar patterns regardless of the scattering theory. All models fall within zones B or C, with no instances reaching the wind-driving threshold of $\Gamma > 1$.
In the inner region, both materials show acceptable gas-to-dust mass ratios (zone C) but insufficient radiation pressure, with $\Gamma$ values typically $\sim$0.01.
In the outer region, Al$_2$O$_3$ models fall into zone B due to excessive depletion requirements, while MgSiO$_3$ models remain in zone C.
These results suggest a two-layer dust structure: Al$_2$O$_3$ can exist in the inner region but cannot explain the observed polarised light in the outer region, while MgSiO$_3$ can be present throughout both regions and accounts for the outer region polarisation.

In contrast, MgFeSiO$_4$ presents a different scenario.
Some models with grain sizes of $\sim$0.056 $\mu$m appear to reach zone D in the inner region, suggesting potential wind-driving capability.
However, these models face two critical limitations: (1) dust temperatures would exceed sublimation thresholds due to strong near-IR absorption \citep{woitke_too_2006}, making their existence impossible in the acceleration zone; and (2) they require gas-to-dust mass ratios at the theoretical minimum for complete silicon depletion, contradicting the $\sim$10\% depletion levels observed \citep{van_de_sande_chemical_2018}.

Scattering theory comparison reveals minimal differences between Mie and DHS approaches.
While Mie theory permits slightly larger maximum grain sizes (0.12 vs 0.09 $\mu$m), this difference is insufficient for any model to reach zone E. 
This demonstrates that grain model uncertainties do not affect our fundamental conclusion about wind-driving insufficiency.

The comprehensive analysis reveals a systematic failure across all dust compositions and scattering theories: while these models successfully reproduce observed polarimetric signatures, no model fits observational constraints while providing sufficient radiation pressure to drive stellar winds in R Dor.

\section{Discussion}
\label{sec:discussion}

\subsection{Two-layer dust envelope structure}
\label{subsubsec:two_layer_structure}

Our zone analysis is consistent with a two-layer dust envelope: an inner gravitationally bound shell containing Al$_2$O$_3$ and an outer unbound envelope dominated by MgSiO$_3$. 
This stratification aligns with previous observations of W Hydrae \citep{khouri_dusty_2015} and theoretical models \citep{hofner_dynamic_2016}. 
Mid-infrared interferometry \citep{karovicova_new_2013} supports this structure in oxygen-rich Mira variables, where alumina grains condense near the stellar surface ($\sim$2R$_{\star}$) and serve as seed particles for silicates forming at larger distances ((4-5)R$_{\star}$). 
Our detection of grains in the inner envelope and their predicted survival only at close distances ($\leq$ 2R$_{\star}$) are consistent with theoretical models of dust-formation zones around AGB stars, where both Al$_2$O$_3$ and MgSiO$_3$ grains could reproduce the observed polarisation signatures.
For the polarised-light observations to be reproduced, we find that silicate grains must exist relatively close to the star $\lesssim 2 R_{\star}$, using as a reference the stellar radius of 27.2 mas in the near-IR reported by \citet{norris_close_2012}.

\subsection{Dust grain size and density profile constraints}
\label{subsubsec:grain_size}

Understanding the optimal dust grain size for AGB wind-driving has evolved from theoretical predictions of $\sim$1 $\mu$m Fe-free silicate grains driving winds through scattering \citep{hofner_winds_2008} to observational evidence of smaller grains $\sim$ 0.3 $\mu$ m contributing to wind acceleration \citep{norris_close_2012}. 
Our R Dor results reveal even smaller grains of $\sim$ 0.1 $\mu$m, consistent with the findings of \cite{khouri_study_2016}. 
This grain size variation appears common across oxygen-rich AGB stars: W Hydrae exhibits larger grains (0.4-0.5 $\mu$m) \citep{ohnaka_clumpy_2016}, while \cite{khouri_inner_2020} report a range from $\geq$ 0.1 $\mu$m (W Hya) to $\leq$ 0.1 $\mu$m (R Crt and SW Vir), suggesting grain growth processes depend on individual stellar characteristics.

Our zone analysis provides a crucial insight: despite confirming $\sim$ 0.1 $\mu$m grains in R Dor, these grains cannot independently drive winds under realistic elemental depletion constraints. 
The analysis considering MgFeSiO$_4$ demonstrates that elemental depletion requirements prevent even iron-bearing silicates from being potential wind drivers. 
These results for R Dor suggest that additional mechanisms must complement dust scattering to produce the observed mass loss.

While our modelling assumes single grain sizes, realistic dust populations follow size distributions.
Our polarimetric observations are most sensitive to grains in the $\sim$ 0.01-0.3 $\mu$m range, where scattering efficiency peaks at visible wavelengths.
Undetected portions of the size distribution are unlikely to significantly alter wind-driving conclusions: larger grains would produce detectable signatures if present in wind-driving quantities, while smaller grains contribute negligibly to radiation pressure.
Most critically, elemental abundance constraints limit the total dust mass regardless of how it is distributed across grain sizes.

We also find a steep dust density profile ($r^{-3.4}$ to $r^{-4.1}$) that aligns with findings for R Dor \citep{khouri_study_2016, ohnaka_clumpy_2016}. 
This steep profile is likely caused by the observed grains being located at the interface between gravitationally bound material and the outflow.
The outward acceleration of the grains can also contribute to the observed steepness of the density profile, but this is probably a minor effect, given that the decrease is approximately 40 times steeper than for a constant-velocity outflow.

\section{Summary and conclusion}
\label{subsec:summary_conclusion}

Our multi-criteria analysis of R Dor indicates that MgSiO$_3$ and Al$_2$O$_3$ grains around this star cannot independently drive its stellar wind under the conditions observed during our study. 
By integrating polarimetric observations with physical constraints of elemental abundances and force balance, we find that even under the most generous assumptions of complete elemental depletion, the radiative force remains insufficient for wind acceleration.

Although Fe-bearing silicates (MgFeSiO$_4$) could theoretically provide adequate radiative pressure in the outer region, two significant challenges prevent them from driving winds: such grains would sublimate in the critical acceleration regions due to overheating \citep{woitke_too_2006}, and they would require unrealistic gas-to-dust mass ratios that lock 80-90\% of available silicon into dust grains - far exceeding the 10\% depletion observed \citep{van_de_sande_chemical_2018}.

While these results represent the first comprehensive test of dust-driven wind theory using observationally constrained models, we emphasise that they apply specifically to R Dor at the time of our observations. 
The wind-driving properties of dust could potentially vary with the stellar pulsation cycle, as shock waves propagating through the circumstellar environment may alter dust grain properties (size, composition, and spatial distribution) and local physical conditions (temperature, density, and velocity structure) that affect radiation pressure efficiency. 
Future multi-epoch observations spanning different pulsation phases would be valuable to assess whether dust-driven wind viability changes throughout the pulsation cycle. 
Our analysis reveals that dust contributes weakly to the momentum budget in this particular case and cannot independently sustain the observed outflow. 

Our results suggest that for R Dor, dust likely plays a supporting rather than dominant role in mass loss, with pulsation-induced processes and other physical mechanisms contributing to the observed outflows. 
This finding motivates expanded studies of other oxygen-rich AGB stars with varying properties to determine whether similar constraints apply more broadly across this stellar population.
Such comprehensive surveys will be essential to assess whether modifications to current mass-loss theories are needed and to develop improved models that capture the full complexity of AGB wind-driving physics.

\begin{acknowledgements}
We thank the referee for their very careful reading of our paper, which has significantly improved its quality. Financial support from the Knut and Alice Wallenberg foundation is gratefully acknowledged through grant no. KAW 2020.0081.
\end{acknowledgements}

\bibliographystyle{aa}
\bibliography{mybib}

\begin{appendix}

\section{PSFs}
\label{sec:PSF}

\begin{figure*}[h]
\centering
\includegraphics[width=\textwidth, trim={0 0cm 0cm 0cm},clip]{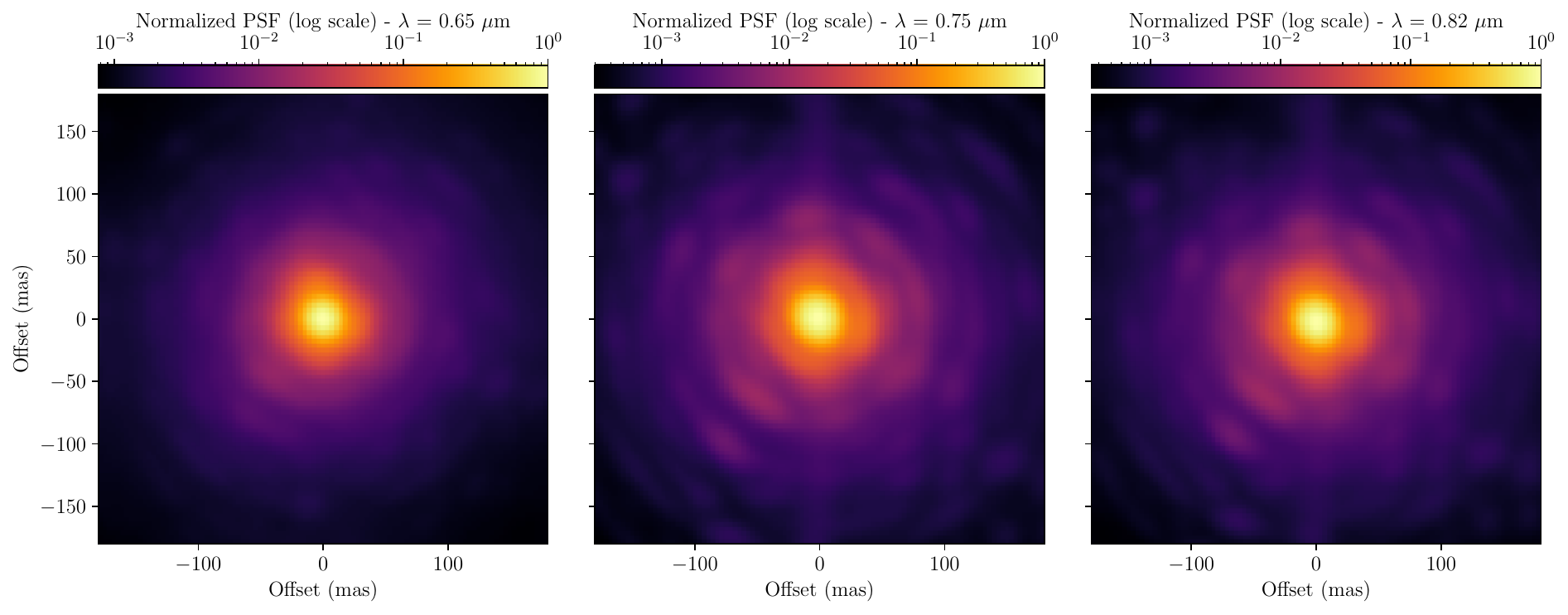}
    \caption{Point Spread Function (PSF) maps for VLT/SPHERE-ZIMPOL observations of R Dor at three wavelengths: 0.65 $\mu$m (left), 0.75 $\mu$m (middle), and 0.82 $\mu$m (right), displayed on a logarithmic scale.}
    \label{fig:PSF_R_Dor}
\end{figure*}

Figure~\ref{fig:PSF_R_Dor} shows the point spread functions obtained from the reference star HD 25170 for the three wavelengths used in our analysis, displayed on a logarithmic scale.

\section{Best-fit models}
\label{sec:best-fit_models}

\subsection{Results for MgSiO$_3$ using DHS}
\label{sec:results_MgSiO3}

Understanding how the six key parameters are constrained is crucial for interpreting the dust properties around R Dor. Table~\ref{tab:dust_parameter_results} summarises the constrained values for the dust grain size ($a_0$), dust density normalisation ($\rho_0$), density profile exponent ($p_\rho$), and the three wavelength-dependent stellar radii ($R_{\lambda_1}$, $R_{\lambda_2}$, $R_{\lambda_3}$). The table includes results for three dust materials (MgSiO$_3$, Al$_2$O$_3$, and MgFeSiO$_4$), each modelled using both Mie theory and the DHS approximation. 

Given the extensive parameter space explored across different materials and scattering theories, we first focus on MgSiO$_3$ using DHS to illustrate how we obtained these constraints. By analyzing the parameter space for this specific case, we show the process that led to the values presented in Table~\ref{tab:dust_parameter_results} before expanding the discussion to other dust compositions and scattering theories.

\begin{figure*}[h]
    \centering
    \includegraphics[width=\textwidth, trim={0cm 0cm 0cm 0cm},clip]{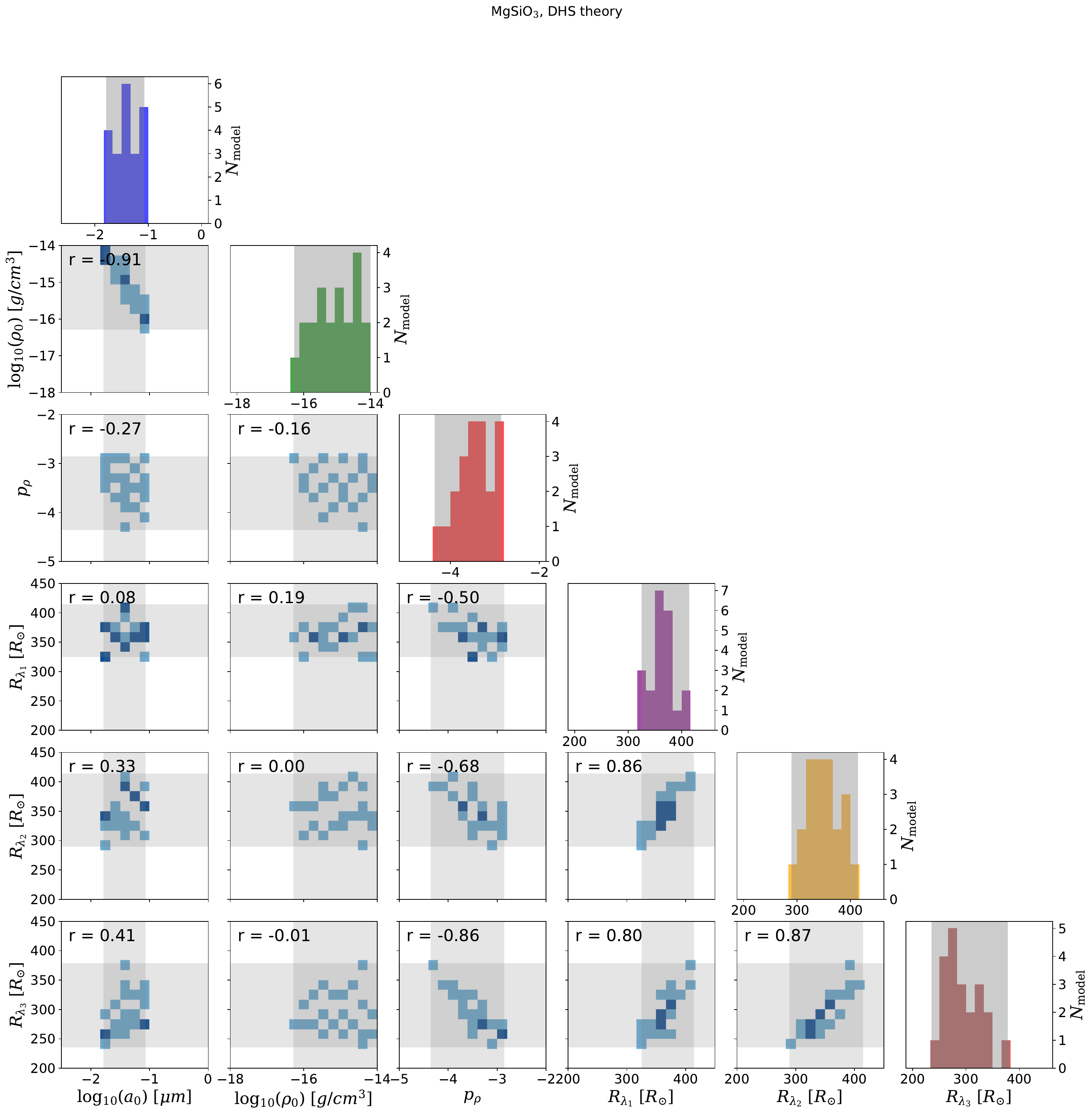}
        \caption{Parameter correlations and distributions for MgSiO$_3$ for the selected models that satisfy the criteria of reduced chi-squared ($\chi^2_{\text{red}} < 1.36$) at the 3$\sigma$ confidence level with a normalised intensity tolerance of $\Delta \mathrm{FW}_{\mathrm{limit}}=\pm30\%$. The diagonal panels show the marginal distributions of the six parameters: $\log_{10}(a_0)$ (dust grain size in microns), $\log_{10}(\rho_0)$ (dust density at $r_0 = 180 \, R_\odot$ in g/cm$^3$), $p_{\rho}$ (density power-law exponent), and stellar radii ($R_{\lambda_1}$, $R_{\lambda_2}$, $R_{\lambda_3}$) in solar radii for the wavelengths 0.65, 0.75, and 0.82 microns, respectively. Vertical and horizontal dashed lines in the histograms represent the median parameter values, and shaded regions indicate parameter ranges. The off-diagonal panels display pairwise scatterplots of the parameters, with denser regions indicating higher model density.}
        \label{fig:chi2min_pyr-mg100_DHS}
    \end{figure*}

Figure~\ref{fig:chi2min_pyr-mg100_DHS} illustrates the parameter space explored in this study for MgSiO$_3$ using DHS, highlighting the distributions and correlations between the six constrained parameters: $\log_{10}(a_0)$, $\log_{10}(\rho_0)$, $p_\rho$, and $R_{\lambda_1}$, $R_{\lambda_2}$, and $R_{\lambda_3}$. The full parameter space consists of 15 bins per parameter, leading to a total of \( 15^6 \) ($1.14 \times 10^7$) computed models. However, only 21 models satisfy the selection criteria described in the previous section, and it is these 21 models that are represented in Figure~\ref{fig:chi2min_pyr-mg100_DHS}.

The diagonal panels in Figure~\ref{fig:chi2min_pyr-mg100_DHS} show the marginal distributions for each parameter, highlighting the minimum and maximum values for models that satisfy our constraints. It is crucial to emphasise that all 21 models represented in these distributions are equally valid solutions that successfully reproduce the observational constraints. For this reason, we focus exclusively on the minimum and maximum parameter values to define the uncertainty ranges, rather than reporting means or standard deviations which would be misleading in this context. Each model within these boundaries represents an equally good fit to the data, making the full range of values physically meaningful. The off-diagonal panels display scatterplots of pairwise parameter relationships, complemented by Spearman correlation coefficients to quantify the strength and direction of these correlations.

\subsubsection{Correlations between parameters} 
\label{subsec:correlation_between_parameters}
The scatter plots in the off-diagonal panels reveal several key correlations and interdependencies between parameters. A notable correlation is observed between $\rho_0$ and $a_0$, where an increase in $a_0$ corresponds to a decrease in $\rho_0$. This correlation reflects a degeneracy between these two parameters: a decrease in $\rho_0$ reduces the total dust cross section, which is compensated by an increase in the grain size, $a_0$. Physically, this suggests that larger grains require lower densities to match the observed scattering and absorption properties, as their interaction with radiation differs from that of smaller grains. Despite this degeneracy, $a_0$ never exceeds an upper limit of approximately 0.09 microns for DHS and 0.12 microns for Mie, regardless of the value of $\rho_0$, providing a strong constraint on the maximum dust grain size. The lower limit on $a_0$, however, depends on the upper limit of $\rho_0$, which we set at $10^{-14} \, \mathrm{g/cm^3}$ based on constraints from the gas density \citep{khouri_empirical_2024} and assuming a very optimistic gas-to-dust ratio of 100. This ratio, typical of the diffuse interstellar medium, is considerably higher than what would be expected in AGB star environments. Therefore, $10^{-14}$ g cm$^{-3}$ represents a genuine physical upper bound for the dust density.
While the lower limit is influenced by the upper limit of $\rho_0$, the key result here is the constraint on the maximum value of $a_0$. 

A strong correlation is also observed between $p_\rho$ and $R_{\lambda_1}$, $R_{\lambda_2}$, and $R_{\lambda_3}$. As $p_\rho$ increases (indicating a less steep density profile), the stellar radii decrease. This relationship likely stems from the interplay between the dust density distribution and the optical depth, which affects the observed stellar radii at different wavelengths. A less steep density profile leads to a more extended dust distribution, reducing the contribution of dust scattering near the star and effectively shifting the apparent stellar radii inward. Finally, there is a strong correlation between $R_{\lambda_1}$, $R_{\lambda_2}$, and $R_{\lambda_3}$: an increase in one implies an increase in the others. This is expected because the stellar radii at different wavelengths are intrinsically linked through the temperature and opacity structure of the stellar atmosphere, and larger radii at shorter wavelengths indicate a globally extended stellar envelope.

\subsubsection{Independent parameter distributions} 
While correlations reveal interdependencies between parameters, examining their independent distributions provides insights into their physically meaningful ranges. For MgSiO$_3$ with DHS scattering, we find the dust size, $a_0$, spans from 0.02 to 0.09 microns. The dust density normalisation, $\rho_0$, spans approximately three orders of magnitude, from $4 \times 10^{-17}$ to $1 \times 10^{-14} \, \mathrm{g/cm^3}$. The power-law exponent governing the dust density profile, $p_\rho$, ranges from $-4.4$ to $-2.8$, indicating a steeper decline than the classical $r^{-2}$ profile expected for steady-state winds with constant outflow velocity.

For the stellar radii, we find consistent patterns across wavelengths. The radius at 0.65 $\mu$m ($R_{\lambda_1}$) ranges from 288 to 414 $R_\odot$, while $R_{\lambda_2}$ (0.75 $\mu$m) spans from 270 to 414 $R_\odot$, and $R_{\lambda_3}$ (0.82 $\mu$m) varies between 216 and 378 $R_\odot$. The pattern $R_{\lambda_1} > R_{\lambda_2} > R_{\lambda_3}$ reflects the wavelength-dependent scattering opacity in the stellar atmosphere, where shorter wavelengths correspond to higher optical depths and larger effective radii. 

Additionally, we confirm that the absolute minima for all parameters fall within the boundaries of our exploration grid, ensuring that the parameter space was well-sampled and that the constraints derived from the models are trustable. This holds true for nearly all parameters, with one notable exception: the dust density parameter ($\rho_0$) consistently reaches its maximum value of $10^{-14}$ g cm$^{-3}$, which corresponds to the upper boundary of our exploration grid. However, as discussed in our analysis of parameter correlations, this limitation primarily affects the minimum value of the grain size parameter ($a_0$) and does not significantly impact our main conclusions.

The polarisation degree panels display a characteristic pattern where polarisation initially increases with radius until reaching a maximum. This behavior occurs because close to the star, most emission originates from the unpolarised stellar surface. As the radius increases, the contribution of polarised scattered light from dust grains becomes more significant until reaching a maximum. Beyond this peak, polarisation degree decreases with increasing radius, corresponding to the decline in dust density, which is primarily responsible for the polarisation effect. Importantly, the observations consistently fall within the range defined by the overlapping models, indicating a reliable fit across all wavelengths. Similarly, the normalised total intensity profiles show excellent agreement between observations and models, exhibiting a smooth decline with radius.

\subsection{Mie vs. DHS}
\label{subsec:mie_vs_dhs}

For MgSiO$_3$, the results obtained using Mie and DHS scattering theories show notable similarities (Table~\ref{tab:dust_parameter_results}). 
Both methods constrain the dust grain size $a_0$ to the same lower limit of 0.02~$\mu$m, with only a marginal difference in upper limits (0.09~$\mu$m for DHS versus 0.12~$\mu$m for Mie) - a distinction that has negligible impact on our conclusions.
This consistency suggests that, for MgSiO$_3$, both scattering theories can reproduce the observed features with similar grain size populations despite their different treatments of grain shapes. 

For the dust density normalisation, $\rho_0$, we observe some differences - $4 \times 10^{-17}$~g~cm$^{-3}$ for the lower bound with DHS versus $10^{-17}$~g~cm$^{-3}$ for MIE - while both share the same upper bound of $10^{-14}$~g~cm$^{-3}$. The power-law exponent $p_\rho$ spans nearly identical ranges for both scattering theories (-4.4 to -2.8), indicating that both approaches can reproduce the observed polarisation patterns with similar dust density distributions. The stellar radii parameters ($R_{\lambda_1}$, $R_{\lambda_2}$, and $R_{\lambda_3}$) also show consistent ranges between the two methods, with only minor differences in the lower bound of $R_{\lambda_1}$ (288 $R_\odot$ for DHS versus 306 $R_\odot$ for Mie). These similarities indicate that both scattering theories can effectively model the observed polarised light using physically comparable dust configurations.

For MgFeSiO$_4$ and Al$_2$O$_3$, we observe similar patterns of consistency between Mie and DHS results. The substantial overlap in parameter ranges across all dust species and both scattering theories suggests that our conclusions about dust properties in R Dor are reliable regardless of which scattering theory is applied.

The inclusion of both Mie and DHS calculations in our analysis serves an important methodological purpose. While Mie theory models perfectly spherical dust grains, the DHS approach approximates irregular grain shapes through a distribution of hollow spheres, offering a more realistic representation of actual cosmic dust particles, which are likely non-spherical in nature. By employing both methods, we test whether our conclusions about dust properties depend on assumptions about grain geometry.

\subsection{Detailed parameter constraints for MgFeSiO$_4$ and Al$_2$O$_3$}

Beyond MgSiO$_3$, we extended our analysis to MgFeSiO$_4$ and Al$_2$O$_3$ (Table~\ref{tab:dust_parameter_results}). The results reveal notable distinctions in the behavior of these dust species.

\begin{figure*}[h]
\centering
\includegraphics[width=\textwidth, trim={0 0cm 0cm 0cm},clip]{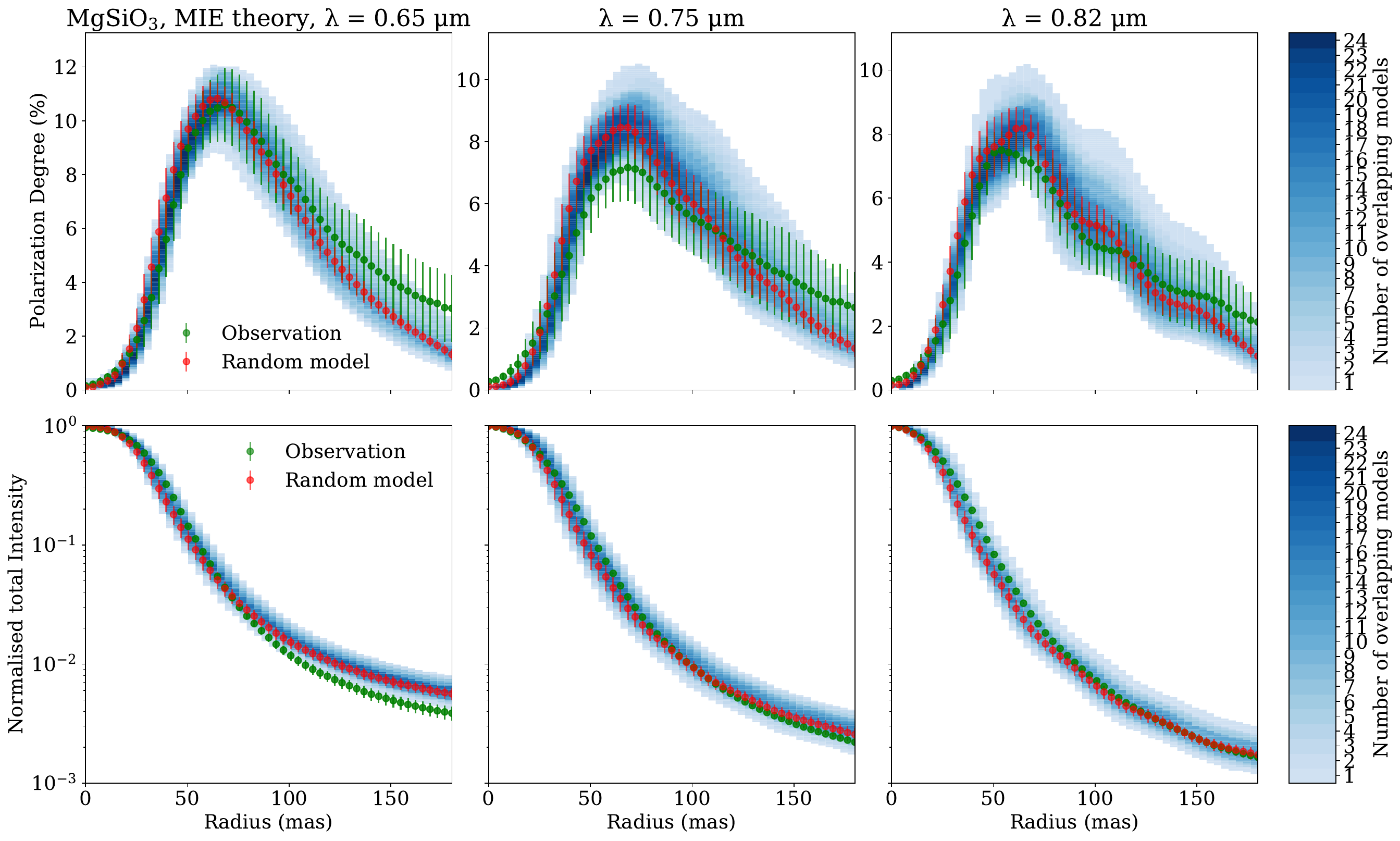}
\includegraphics[width=\textwidth, trim={0 0cm 0cm 0cm},clip]{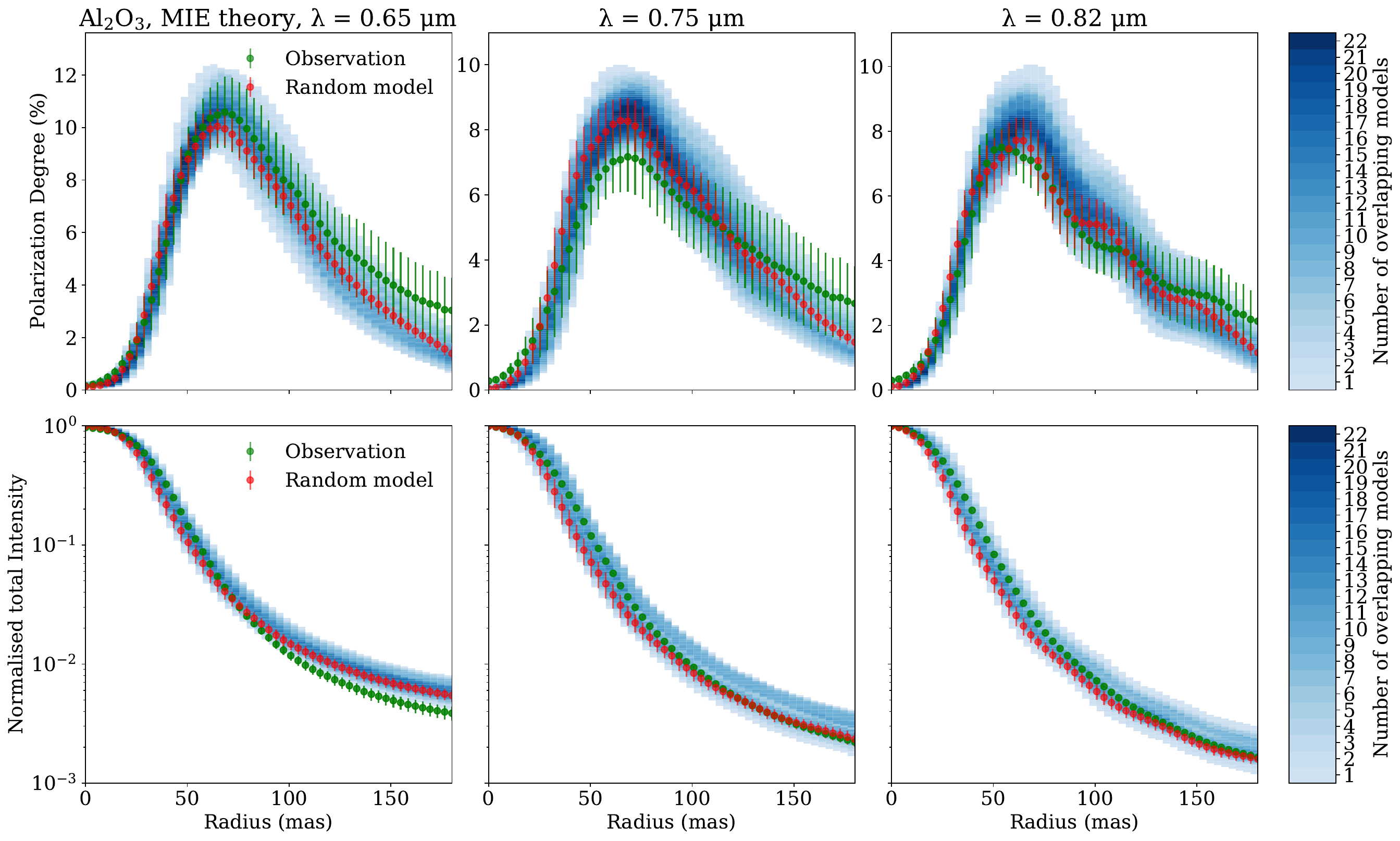}
    \caption{Same as Fig.\,\ref{fig:PD_Itot_all_models_pyr-mg100_DHS} but for MgSiO\(_3\) and using the Mie theory (top) and for Al\(_2\)O\(_3\) and using the Mie theory (bottom).}
    \label{fig:PD_Itot_all_models_pyr-mg100_cor-c_MIE}
\end{figure*}

\begin{figure*}[h]
\centering
\includegraphics[width=\textwidth, trim={0 0cm 0cm 0cm},clip]{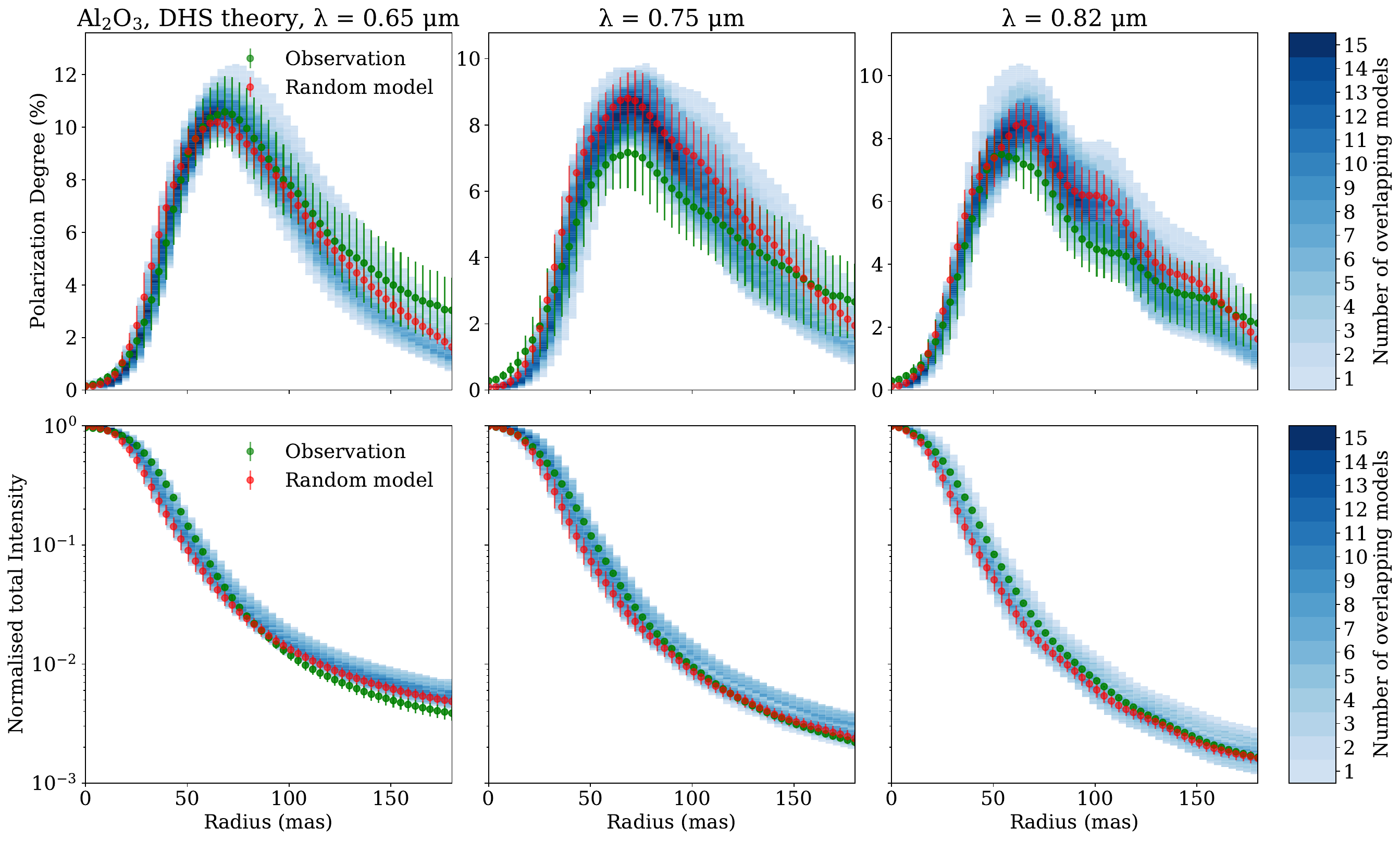}
\includegraphics[width=\textwidth, trim={0 0cm 0cm 0cm},clip]{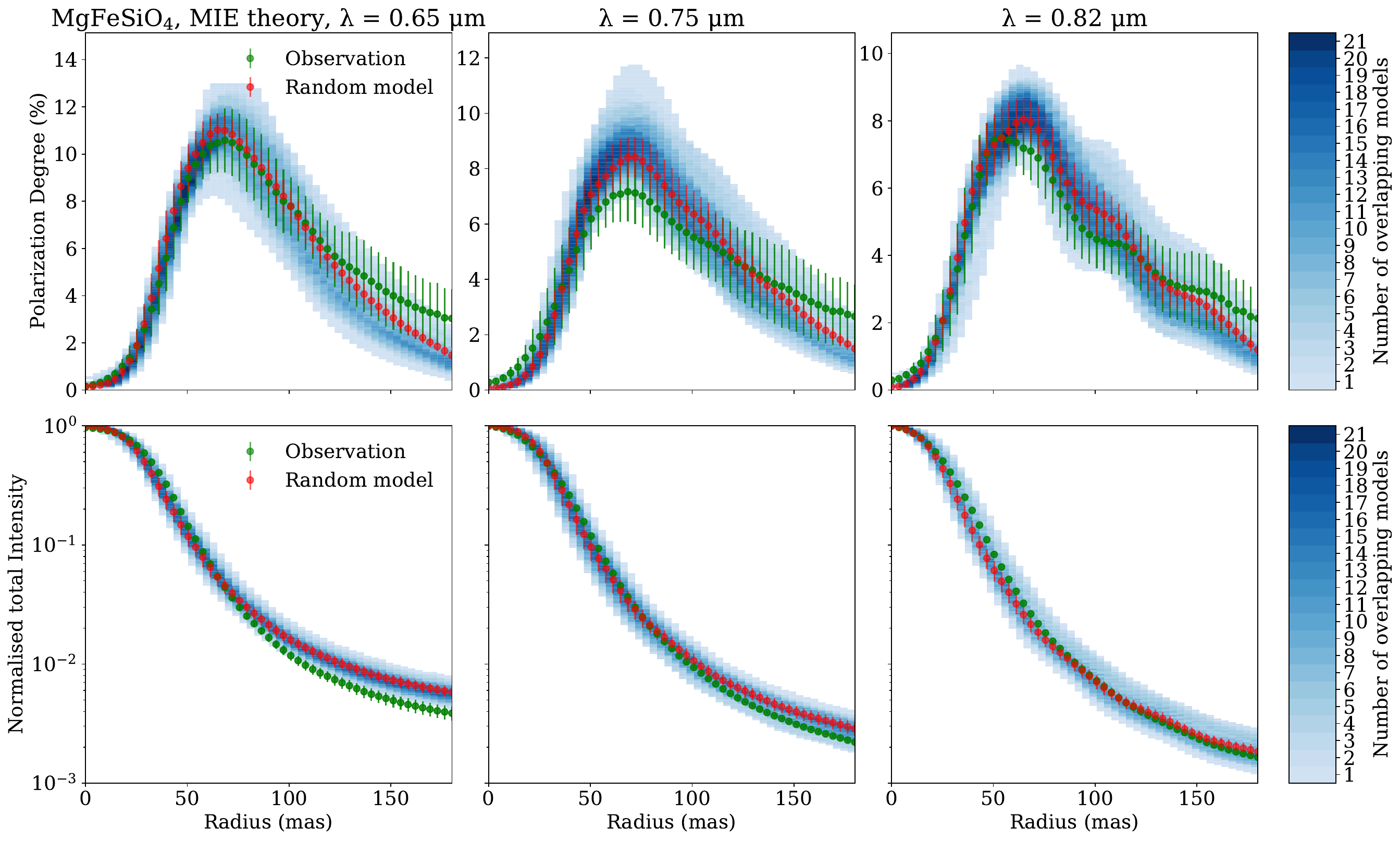}
    \caption{Same as Fig.\,\ref{fig:PD_Itot_all_models_pyr-mg100_DHS} but for Al\(_2\)O\(_3\) and using the DHS theory (top) and for MgFeSiO\(_4\) and using the Mie theory (bottom).}
    \label{fig:PD_Itot_all_models_cor-c_DHS_ol-mg50_MIE}
\end{figure*}

\begin{figure*}[h]
\centering
\includegraphics[width=\textwidth, trim={0 0cm 0cm 0cm},clip]{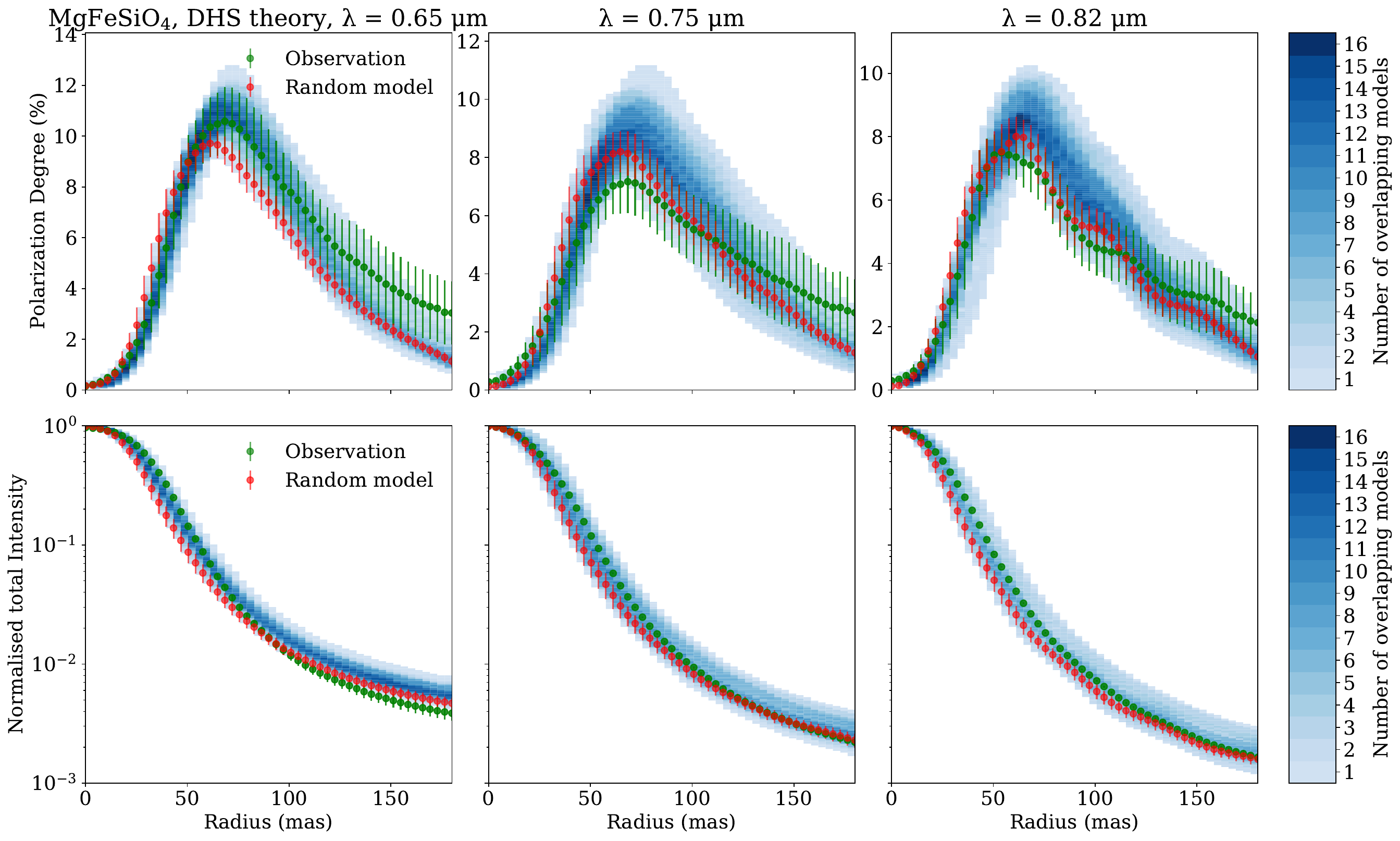}
\caption{Same as Fig.\,\ref{fig:PD_Itot_all_models_pyr-mg100_DHS} but for MgFeSiO\(_4\) and using the DHS theory.}
\label{fig:PD_Itot_all_models_ol-mg50_DHS}
\end{figure*}

The comparison between the modelled and observed polarisation degree and total intensity profiles is presented in Fig.~\ref{fig:PD_Itot_all_models_pyr-mg100_cor-c_MIE} for MgSiO$_3$ using Mie, in Fig.~\ref{fig:PD_Itot_all_models_cor-c_DHS_ol-mg50_MIE} for Al$_2$O$_3$ using DHS and MgFeSiO$_4$ using Mie, and in Fig.~\ref{fig:PD_Itot_all_models_ol-mg50_DHS} for MgFeSiO$_4$ using DHS.

For MgFeSiO$_4$, the grain size parameter $a_0$ is constrained to 0.06 to 0.09 $\mu$m for DHS and 0.06 to 0.12 $\mu$m for Mie theory, as shown in Table~\ref{tab:dust_parameter_results}. Notably, the minimum grain size for MgFeSiO$_4$ is larger than for MgSiO$_3$ (0.02 $\mu$m) across both scattering theories. The dust density normalisation $\rho_0$ for MgFeSiO$_4$ shows identical ranges for both scattering theories: $2 \times 10^{-17}$ to $10^{-14}$ g/cm$^3$. The power-law exponent $p_\rho$ ranges from -4.8 to -3.0 for DHS and -4.4 to -3.0 for Mie, indicating a slightly steeper density gradient is possible when modelling with DHS for this material.

For Al$_2$O$_3$, the grain size parameter $a_0$ is constrained to 0.04 to 0.09 $\mu$m for DHS and 0.04 to 0.12 $\mu$m for Mie. The dust density normalisation $\rho_0$ for Al$_2$O$_3$ reaches the lowest values among the three materials: $6.3 \times 10^{-18}$ to $10^{-14}$ g/cm$^3$ for DHS and $10^{-17}$ to $10^{-14}$ g/cm$^3$ for Mie. The power-law exponent $p_\rho$ for Al$_2$O$_3$ spans -4.8 to -3.0 for both scattering theories. The stellar radii for Al$_2$O$_3$ show ranges of 306-414 $R_\odot$ for $R_{\lambda_1}$ with DHS (306-432 $R_\odot$ with Mie), 216-396 $R_\odot$ for $R_{\lambda_2}$ with DHS (198-414 $R_\odot$ with Mie), and 198-360 $R_\odot$ for $R_{\lambda_3}$ with both theories.

For both Al$_2$O$_3$ and MgFeSiO$_4$, the stellar radii parameters ($R_{\lambda_1}$, $R_{\lambda_2}$, and $R_{\lambda_3}$) show consistent wavelength-dependent patterns across all materials. MgFeSiO$_4$ models exhibit a somewhat broader range of stellar radii values, particularly at the shortest wavelength (0.65 $\mu$m), where DHS modelling yields radii as low as 234 $R_\odot$ compared to the 306 $R_\odot$ lower bound found with Mie. This difference may reflect how the scattering and absorption properties of iron-bearing silicates interact with stellar radiation differently than iron-free materials. As shown in Table~\ref{tab:dust_parameter_results}, the stellar radii for MgFeSiO$_4$ range from 234-432 $R_\odot$ for $R_{\lambda_1}$ with DHS (306-432 $R_\odot$ with Mie), 198-432 $R_\odot$ for $R_{\lambda_2}$ with both theories, and 198-360 $R_\odot$ for $R_{\lambda_3}$ with both theories, consistent with the values presented in our parameter exploration.

The overall consistency in parameter ranges between Mie and DHS results, despite some slight variations in upper and lower bounds, suggests that our conclusions about dust properties in R Dor are valid across different assumptions about grain geometry. A key finding across all three dust materials is the convergence on similar grain size constraints, with upper limits consistently around 0.1 $\mu$m regardless of the scattering theory employed or the dust composition. Equally significant is the power-law exponent $p_\rho$, which ranges from -4.8 to -2.8 across all materials - markedly steeper than the $r^{-2}$ profile typically expected in steady-state wind models.

\section{Elemental-abundance constraints}
\label{sec:elemental-abundance-constraints}

When modelling dust-driven winds in AGB stars, it is essential to consider the elemental abundances available for dust formation. 
The total abundance of elements (both in gas phase and locked in dust grains) around R Dor is expected to be similar to solar abundances \citep{asplund_chemical_2009}. 
This places fundamental constraints on the maximum amount of dust that can form from the available gas, and consequently on the minimum gas-to-dust mass ratio that is physically realistic.

For the three dust species considered in this study, the limiting elements are silicon (Si) for silicate grains (MgSiO$_3$ and MgFeSiO$_4$) and aluminium (Al) for alumina (Al$_2$O$_3$). 
These elements have the lowest abundances relative to their stoichiometric requirements in the respective dust compounds, making them the bottleneck for dust formation.

The gas-to-dust mass mass ratio limit can be derived from the depletion fraction of the limiting element. For a given dust species, this relationship is expressed as:

\begin{equation}
\left(\frac{\rho_{\mathrm{g}}}{\rho_{\mathrm{d}}}\right)_{\mathrm{min}} = \frac{1}{X_{\mathrm{dust}}} = \frac{1}{f_{\mathrm{X}} \times A_{\mathrm{X}} \times \frac{m_{\mathrm{dust}}}{m_{\mathrm{X}}}},
\end{equation}

where $X_{\mathrm{dust}}$ is the dust mass fraction, $f_{\mathrm{X}}$ is the fraction of the limiting element X (Si or Al) depleted from the gas phase into dust grains, $A_{\mathrm{X}}$ is the abundance of element X relative to hydrogen by mass (based on solar abundances), and $\frac{m_{\mathrm{dust}}}{m_{\mathrm{X}}}$ is the ratio of the total dust mass to the mass of element X in the dust compound.

Using the solar photospheric abundances from \citet{asplund_chemical_2009}, where $\log(N_{\mathrm{Si}}/N_{\mathrm{H}}) + 12 = 7.51$ and $\log(N_{\mathrm{Al}}/N_{\mathrm{H}}) + 12 = 6.45$, we can calculate the corresponding mass fractions relative to the total gas mass. This conversion involves several steps:

First, we convert the logarithmic abundance notation to linear number ratios:
\begin{align*}
\frac{N_{\mathrm{Si}}}{N_{\mathrm{H}}} &= 10^{(7.51-12)} \approx 3.24 \times 10^{-5} \\
\frac{N_{\mathrm{Al}}}{N_{\mathrm{H}}} &= 10^{(6.45-12)} \approx 2.82 \times 10^{-6}
\end{align*}

Next, we convert these number ratios to mass ratios relative to hydrogen by multiplying by the ratio of atomic masses:
\begin{align*}
\frac{m_{\mathrm{Si}}}{m_{\mathrm{H}}} &= \frac{N_{\mathrm{Si}}}{N_{\mathrm{H}}} \times \frac{M_{\mathrm{Si}}}{M_{\mathrm{H}}} \approx 3.24 \times 10^{-5} \times \frac{28.09}{1.008} \approx 9.03 \times 10^{-4} \\
\frac{m_{\mathrm{Al}}}{m_{\mathrm{H}}} &= \frac{N_{\mathrm{Al}}}{N_{\mathrm{H}}} \times \frac{M_{\mathrm{Al}}}{M_{\mathrm{H}}} \approx 2.82 \times 10^{-6} \times \frac{26.98}{1.008} \approx 7.55 \times 10^{-5}
\end{align*}
where $M_{\mathrm{Si}} = 28.09$ amu, $M_{\mathrm{Al}} = 26.98$ amu, and $M_{\mathrm{H}} = 1.008$ amu are the atomic weights.

Finally, we account for the fact that hydrogen constitutes exactly 73.81\% of the total gas mass in the solar composition, with the remainder consisting of helium (24.85\%) and heavier elements (1.34\%) \citep{asplund_chemical_2009}. The mass fractions relative to the total gas mass are therefore:
\begin{align*}
A_{\mathrm{Si}} &= \frac{m_{\mathrm{Si}}}{m_{\mathrm{total}}} = \frac{m_{\mathrm{Si}}}{m_{\mathrm{H}}} \times \frac{m_{\mathrm{H}}}{m_{\mathrm{total}}} \approx 9.03 \times 10^{-4} \times 0.7381 \approx 6.7 \times 10^{-4} \\
A_{\mathrm{Al}} &= \frac{m_{\mathrm{Al}}}{m_{\mathrm{total}}} = \frac{m_{\mathrm{Al}}}{m_{\mathrm{H}}} \times \frac{m_{\mathrm{H}}}{m_{\mathrm{total}}} \approx 7.55 \times 10^{-5} \times 0.7381 \approx 5.6 \times 10^{-5}
\end{align*}

For MgSiO$_3$ (pyroxene), the mass fraction of silicon can be calculated from the atomic weights:
\begin{align*}
\frac{m_{\mathrm{Si}}}{m_{\mathrm{MgSiO_3}}} &= \frac{28.09}{24.31 + 28.09 + 3 \times 16.00} \\
&= \frac{28.09}{100.40} \approx 0.28 \text{ or } 28\%
\end{align*}

For MgFeSiO$_4$ (olivine), the mass fraction of silicon is:
\begin{align*}
\frac{m_{\mathrm{Si}}}{m_{\mathrm{MgFeSiO_4}}} &= \frac{28.09}{24.31 + 55.85 + 28.09 + 4 \times 16.00} \\
&= \frac{28.09}{172.25} \approx 0.163 \text{ or } 16.3\%
\end{align*}

For Al$_2$O$_3$ (alumina), the mass fraction of aluminum is:
\begin{align*}
\frac{m_{\mathrm{Al}}}{m_{\mathrm{Al_2O_3}}} &= \frac{2 \times 26.98}{2 \times 26.98 + 3 \times 16.00} \\
&= \frac{53.96}{101.96} \approx 0.53 \text{ or } 53\%
\end{align*}

Assuming complete depletion of the limiting element ($f_{\mathrm{X}} = 1$), the minimum gas-to-dust mass ratios would be approximately 420 for MgSiO$_3$, 240 for MgFeSiO$_4$, and 9400 for Al$_2$O$_3$.

These constraints on the gas-to-dust mass ratio are incorporated into our assessment of wind-driving potential. Models that predict gas-to-dust mass ratios below these minimum values would require unrealistic levels of element depletion and are therefore excluded from consideration, regardless of their ability to fit the observed polarisation data.

Substituting the calculated values into our equation, we can express the minimum gas-to-dust mass ratio as a function of the depletion fraction for each dust species:

\begin{align}
\left(\frac{\rho_{\mathrm{g}}}{\rho_{\mathrm{d}}}\right)_{\mathrm{min, MgSiO_3}} &\approx \frac{420}{f_{\mathrm{Si}}} \label{eq:min_ratio_MgSiO3} \\
\left(\frac{\rho_{\mathrm{g}}}{\rho_{\mathrm{d}}}\right)_{\mathrm{min, MgFeSiO_4}} &\approx \frac{240}{f_{\mathrm{Si}}} \label{eq:min_ratio_MgFeSiO4} \\
\left(\frac{\rho_{\mathrm{g}}}{\rho_{\mathrm{d}}}\right)_{\mathrm{min, Al_2O_3}} &\approx \frac{9400}{f_{\mathrm{Al}}} \label{eq:min_ratio_Al2O3}
\end{align}

These expressions provide a direct way to estimate the minimum physically plausible gas-to-dust mass ratio for a given depletion fraction of the limiting element in each dust species. 

Observations of gas-phase abundances in AGB stars suggest that depletion is typically incomplete. For R Dor specifically, we can estimate the silicon depletion fraction using the SiO abundance measurements from \citet{van_de_sande_chemical_2018}. Their observations indicate that the SiO abundance relative to H$_2$ in the circumstellar envelope of R Dor is between $5.5 \times 10^{-5}$ and $6.0 \times 10^{-5}$. Comparing this to the expected total silicon abundance (relative to H$_2$) of approximately $6.5 \times 10^{-5}$ based on solar values \citep{asplund_chemical_2009}, we can calculate the fraction of silicon depleted into dust grains as follows:

\begin{align}
f_{\mathrm{Si}} &= \frac{[\mathrm{Si}]_{\mathrm{total}} - [\mathrm{Si}]_{\mathrm{gas}}}{[\mathrm{Si}]_{\mathrm{total}}} \\
&= \frac{[\mathrm{Si}]_{\mathrm{total}} - [\mathrm{SiO}]_{\mathrm{gas}}}{[\mathrm{Si}]_{\mathrm{total}}} \\
&= \frac{6.5 \times 10^{-5} - (5.5 \text{ to } 6.0) \times 10^{-5}}{6.5 \times 10^{-5}} \\
&= \frac{(0.5 \text{ to } 1.0) \times 10^{-5}}{6.5 \times 10^{-5}} \\
&\approx 0.08 \text{ to } 0.15
\end{align}

This calculation shows that only about 8-15\% of the available silicon is depleted into dust grains. Substituting these depletion fractions ($f_{\mathrm{Si}} = 0.08$ and $0.15$) into Equations~\eqref{eq:min_ratio_MgSiO3} and~\eqref{eq:min_ratio_MgFeSiO4} yields minimum gas-to-dust mass ratios of approximately 2800-5300 for MgSiO$_3$ and 1600-3000 for MgFeSiO$_4$.

To assess whether our dust models are physically plausible, we compare the gas-to-dust mass ratios derived from our polarimetric observations with the minimum values calculated above. Our primary criterion is that the gas-to-dust mass ratio should exceed the minimum theoretical value derived from elemental abundances. This minimum threshold corresponds to 100\% of the limiting element (silicon or aluminum) being locked into dust grains - a scenario that represents an absolute physical limit rather than a likely reality. We deliberately adopt this flexible criterion to give our dust models the maximum opportunity to satisfy physical constraints. Using this threshold allows us to identify which dust models are physically impossible (those requiring more than 100\% of available elements) versus those that are at least theoretically possible, if perhaps implausible. When comparing our models to the more realistic depletion fractions derived from gas-phase observations (approximately 10\% for silicon according to \citet{van_de_sande_chemical_2018}), we can further distinguish between models that are merely possible and those that align with actual circumstellar conditions.

\section{Force balance and $\Gamma$ calculation}
\label{sec:force_balance}

Dust-driven winds rely on the balance between stellar radiation pressure and gravitational forces acting on dust grains. To assess the wind-driving potential of the dust with the grain sizes, densities, and compositions derived in this study, we evaluate the force balance parameter $\Gamma$ - the ratio of radiative pressure to gravitational force. Following \citet{hofner_mass_2018}, we use an expression for $\Gamma$ that accounts for wind dynamics by incorporating gas and dust velocities:

\begin{equation}
\label{eq:gamma}
\Gamma = \frac{\langle \kappa_{\mathrm{pr}} \rangle L_{\star}}{4 \pi c G M_{\star}} \frac{v_{\mathrm{g}}}{r_{\mathrm{gd}}v_{\mathrm{d}}},
\end{equation}

where $L_{\star}$ is the stellar luminosity, $c$ is the speed of light, $G$ is the gravitational constant, $M_{\star}$ is the stellar mass, $v_{\mathrm{g}}$ is the gas velocity, $v_{\mathrm{d}}$ is the dust velocity, $r_{\mathrm{gd}}$ represents the gas-to-dust mass ratio (defined as $\rho_{\mathrm{g}}/\rho_{\mathrm{d}}$), and $\langle \kappa_{\mathrm{pr}} \rangle$ is the flux-mean radiative pressure opacity per unit mass of dust. This flux-mean opacity is defined as:

\begin{equation}
\langle \kappa_{\mathrm{pr}} \rangle = \frac{\int \kappa_{\mathrm{pr}}(\lambda, a) \, B_{\lambda}(T_{\star}) \, d\lambda}{\int B_{\lambda}(T_{\star}) \, d\lambda}.
\end{equation}

Here, $\kappa_{\mathrm{pr}}(\lambda, a)$ is the radiative pressure cross-section at wavelength $\lambda$, and $B_{\lambda}(T_{\star})$ is the Planck function at stellar temperature $T_{\star}$. The radiative pressure cross-section accounts for both absorption and scattering contributions:

\begin{equation}
\kappa_{\mathrm{pr}}(\lambda, a) = \kappa_{\mathrm{abs}}(\lambda, a) + (1 - g) \, \kappa_{\mathrm{sca}}(\lambda, a),
\end{equation}

where $\kappa_{\mathrm{abs}}$ and $\kappa_{\mathrm{sca}}$ are the absorption and scattering cross-sections, respectively, and $g$ is the scattering asymmetry parameter.

\section{Additional figures}
\label{sec:additional_figures}

\end{appendix}
\end{document}